\documentclass[12pt]{JHEP3}
\usepackage{amsmath,amsfonts,times}

\newcommand{\bea}{\begin{eqnarray}}

\newcommand{\eea}{\end{eqnarray}}

\newcommand{\lala}{\langle\!\langle}
\newcommand{\rr}{\rangle\!\rangle}

\newcommand{\sst}{\scriptscriptstyle}

\newcommand{\1}{\one}
\newcommand{\2}{\two}

\newcommand{\beq}{\begin{equation}}
\newcommand{\eeq}{\end{equation}}

\newcommand{\rf}[1]{(\ref{#1})}

\newcommand{\pa}{\partial}
\newcommand{\ot}{\otimes}
\newcommand{\ra}{\rightarrow}

\newcommand{\ti}{\times}

\newcommand{\fr}[2]{{\textstyle \frac{#1}{#2} }}

\newcommand{\df}{\equiv}

\newcommand{\Ga}{\Gamma}
\newcommand{\de}{\delta}

\newcommand{\ep}{\epsilon}
\newcommand{\la}{\lambda}
\newcommand{\om}{\omega}

\newcommand{\si}{\sigma}

\newcommand{\vf}{\varphi}

\newcommand{\bfc}{{\rm\mathbf c}}
\newcommand{\bfd}{{\rm\mathbf d}}
\newcommand{\rma}{{\rm a}}
\newcommand{\rmb}{{\rm b}}
\newcommand{\rmc}{{\rm c}}
\newcommand{\rmd}{{\rm d}}

\newcommand{\CF}{{\mathcal F}}

\newcommand{\CH}{{\mathcal H}}

\newcommand{\CK}{{\mathcal K}}

\newcommand{\CM}{{\mathcal M}}

\newcommand{\CO}{{\mathcal O}}

\newcommand{\CU}{{\mathcal U}}

\newcommand{\CZ}{{\mathcal Z}}

\newcommand{\SH}{{\mathsf H}}

\newcommand{\SM}{{\mathsf M}}

\newcommand{\SP}{{\mathsf P}}

\renewcommand{\sf}{{\mathsf f}}
\newcommand{\sh}{{\mathsf h}}

\newcommand{\su}{{\mathsf u}}

\newcommand{\0}{{\mathfrak 0}}
\newcommand{\one}{{\mathfrak 1}}
\newcommand{\two}{{\mathfrak 2}}

\newcommand{\BN}{{\mathbb N}}

\newcommand{\BR}{{\mathbb R}}

\newcommand{\BZ}{{\mathbb Z}}

\title{On tachyon condensation and
open-closed duality in the
$c=1$ string theory}

\author{J\"org Teschner
\vspace{5mm}
\\
{\it Institut f\"ur theoretische Physik
\\ Freie Universit\"at Berlin
\\ Arnimallee 14, 14195 Berlin, Germany \vspace{5mm}\\
Max Planck-Institut f\"ur Gravitationsphysik
\\ Am M\"uhlenberg 1
\\ 14476 Golm, Germany}
\\ {\tt teschner@physik.fu-berlin.de }
}

\abstract{We present an exact representation
for decaying ZZ-branes within the dual matrix model
formulation of $c=1$ string theory. It is shown how
to trade the insertion of decaying ZZ-branes
for a shift of the closed string background. Our
formalism allows us to demonstrate that the
conjectured world-sheet mechanism behind the
open-closed dualities (summing over disc insertions)
is realized here in a clear way. On the way we
need to clarify certain infrared issues -
insertion of ZZ-branes creates solitonic
superselection sectors.
}

\preprint{
\hepth{0504043} \\
 }

\begin{document}

%\begin{frontmatter}
%\title{On open-closed duality in the
%$c=1$ string theory}
%\author[SpbU]{JT\corauthref{cor1}}
%\ead{ff}
%\address[SpbU]{Berlin}
%\corauth[cor1]{Corresponding author}
%\date{}
%\begin{abstract}
\abstract{
We study the decay of the unstable $D0$-branes
in the $c=1$ noncritical string theory
with the help of the duality with free fermionic
field theory.
%\end{abstract}
}
%\begin{keyword}D-brane, tachyon, bosonization
%\PACS 11.25.Hf; 11.30.Pb; 02.20.Uw; 02.20.Tw
%\end{keyword}
%\end{frontmatter}

\section{Introduction}

The study of noncritical string theories has recently seen a
renaissance, initiated by the appearance of \cite{McGV,KMS,MTV}.
One of the reasons for the renewed interest in these string
theoretical toy models is the observation that the dualities
between noncritical string theories and matrix models can be seen
as examples for open-closed dualities in string theory. In
particular it was proposed in \cite{McGV,KMS} that the free
fermionic field theory conjectured to be dual to the $c=1$ string
theory is nothing but the open string theory on a ``gas'' of
unstable D0 branes. This was recently further substantiated in
\cite{TT}. Another example in which the mechanism of open-closed
dualities is exhibited in a particularly explicit way is the
duality between the Kontsevich matrix model and pure topological
gravity \cite{GR}.

Another source of recent interest in these toy models was the realization
that nonperturbatively stable definitions of the relevant theories
exist \cite{TT,DKKMMS}, after all \cite{Pol}.
These developments open the possibility to study
certain time-dependent phenomena such as D-brane decay
in an exactly soluble framework. One may therefore
hope to improve our understanding of certain
foundational aspects of string theory for which
time-dependent phenomena represent a challenge.
One may in particular hope to learn how to describe the
final state of a decaying D-brane, and to what extend
one may describe the process with the help of the
usual perturbative approach to string theory.

Our initial aim was therefore to use the duality between
noncritical string theories and matrix models in order to
find an exact description for the decay of unstable
D0-branes in the two-dimensional string theory.
It then becomes possible to learn about scope and
limitations of the perturbative approach to the same
problem.

A particularly interesting feature of the exact description
for the D-brane decay in two-dimensional string theory that
we are about to present is the fact that it exhibits
an example of open-closed duality in a very explicit way:
Insertion of decaying D0-branes can be traded for a shift
of the closed string background. It is furthermore
possible to show that this shift of the background is
perturbatively generated by summing over the disc insertions
which represent the emission of closed strings from the
decaying brane. This fits well into the picture
proposed in \cite{McGV,KMS}: If the $c=1$ background is
generated by the insertion of a gas unstable D0 branes,
it should be possible to trade addition of further
probe D-branes for a shift of the closed string
background.

On  the way we will need to clarify certain
infrared issues. It will be shown that the
insertion of D-branes creates ``solitonic'' superselection
sectors. D-branes are solitons after all. Excitations
in these sectors can not be represented by normalizable
vectors in the sector which describes pure closed string
excitations. Nevertheless there is a clear sense in
which these sectors are equivalent to the sector
with zero D-branes: These sectors can not be
distinguished by measuring any local observable like
the expectation values of the tachyon field. They are
distinguished by the values of {\it global} observables,
though.
This limits the extend to which a narrow-minded
version of open-closed
duality is true: The trade of D-branes for a shift of the
closed string background is not perfect, it works
to the extend to which we may regard the different
sectors as physically equivalent.
For some questions it may nevertheless be
important to keep in mind
that the insertion of D0-branes does not generate
a {\it normalizable} deformation of the background,
similar to the phenomenon emphasized in \cite{SS}.

Previous work on similar questions is contained in
\cite{GK,BSVY,AJ}. The present paper will describe a new approach
to this problem which allows us to go somewhat further and to
clarify a number of aspects which have not yet been discussed in
the literature. In order to simplify the presentation we have
chosen to focus of the case of the bosonic $c=1$ string theory.
However, a good part of our formalism carries over with only small
changes to the case of type 0B $\hat{c}=1$ string theory.

\section{The $c=1$ string as free fermionic field theory}

We are going to revisit some aspects of the
conjectured duality between the $c=1$ string theory and free
fermionic field theory, see \cite{GM} for a review.
One of our main aims is to introduce a formalism which will
be particularly well-suited for our later discussions of
D-branes within the free fermionic field theory.
This will also allow us to present
a simplified representation for the S-matrix
of bosonic excitations \cite{M,MPR} within the free fermionic
field theory.

The presentation will be brief, the necessary technical
details are contained in the appendix A.

\subsection{$c=1$ string}

The $c=1$ string theory is a two-dimensional string background
with coordinates $(X_0,\phi)\in\BR^2$, where $X_0$ represents
time. This background is characterized by the following
expectation values for the target-space metric $G_{\mu\nu}$,  the
dilaton $\Phi$ and the tachyon field $T$:
\begin{equation}
G_{\mu\nu}=\eta_{\mu\nu},\quad \Phi=\phi,\quad T=\mu e^{2\phi}\,.
\end{equation}
The worldsheet-description of this theory  % \cite{DFK}\cite{DKKMMS}
is characterized by the world-sheet action
\begin{equation}\label{WSact}
S=\int d^2 x \;
\bigg(-\frac{1}{4\pi}\pa_+ X_0\pa_-
X_0+\frac{1}{\pi}\pa_+\phi
\pa_-\phi
-\mu(2\phi+\ln\pi\mu) e^{2\phi}\bigg)+({\rm ghosts})\, .
\end{equation}

The string theory has one propagating space-time field, the
tachyon $T$ . The vertex operators which create the modes of this
field with definite space-time energy $\omega$ will be denoted as
$T_{\iota}(\omega)$, where $\iota=-$  creates the asymptotic
in-states, whereas $\iota=+$ corresponds to the asymptotic
out-states,
\begin{equation}
T_{\pm}(\omega)\,=\,e^{-i\om X_0}\,e^{2(1\mp i\om)\phi}.
  \end{equation}
Standard CFT methods will allow us to
define arbitrary string scattering amplitudes
in an asymptotic
expansion in powers of the string coupling constant
$g_s$,
\begin{equation}\begin{aligned}
\big\langle\,T_{\rm\sst out}({\omega}_1) & \dots
T_{\rm\sst out}({\omega}_n)\,
T_{\rm\sst in}({\omega}_1')\dots
T_{\rm\sst in}({\omega}_m')\,
\big\rangle_{c=1}^{}
\,\equiv\,\\ & \,\equiv\,
\sum_{h=0}^{\infty}\,
g_s^{2h-2}\,\big\langle\,T_{\rm\sst out}({\omega}_1)\dots
T_{\rm\sst out}({\omega}_n)\,
T_{\rm\sst in}({\omega}_1')\dots
T_{\rm\sst in}({\omega}_m')\,
\big\rangle^{\sst(h)}_{{c=1},}
\end{aligned}\end{equation}
The amplitudes $\langle\cdots\rangle^{\sst(h)}_{\sst c=1}$
associated to Riemann surfaces with genus $h$ are defined in the
usual way by integrating CFT-correlation functions over the moduli
space of Riemann surfaces.

\subsection{Free fermionic field theory}

Let us consider the quantum field theory of free fermions
in the inverted
harmonic oscillator
potential. The one-particle Hamiltonian will be
\begin{equation} \textsf{h} = -
\frac{\rmd^2}{\rmd \lambda^2} -\frac{1}{4} \lambda^2 \label{1ptcH}.  \end{equation}
There exists a complete set of {\it real} eigenfunctions
\[
\big\{\,{\rm F}_p(\omega|\lambda)\,;\,\omega\in\BR\,,\,p\in\{+,-\}\,
\big\}\,
\]
such that the labels $(\om,p)$ of ${\rm F}_p(\omega|.)$ correspond
to the eigenvalues of $\sh$ and the parity operator $\SP$
respectively.

For each pair $(\omega,p)$ of eigenvalues for the Hamiltonian
\textsf{h} and parity \textsf{P}, we introduce a pair of fermionic
creation- and annihilation operators $(\rmc^{\dagger}_p(\om),
\rmc_p^{\phantom{\dagger}}(\om))$ and require that they satisfy
the canonical anticommutation relations
\begin{equation}
\big\{~\rmc_{p_\1}^{\phantom{\dagger}}(\om)~,~
\rmc^{\dagger}_{p_\2}(\om')~\big\}\;=\;
\de_{p_\1p_\2}\de(\om-\om').
\end{equation}
We shall also use the vector notation
\[
{\bfc}(\om)=\bigg(\begin{matrix}
{\rmc}_+(\om)\\ % [-1ex]
{\rmc}_-(\om)
\end{matrix}\bigg),\qquad
{\rm\mathbf F}(\om|\la)=\bigg(\begin{matrix} {\rm F}_+(\om|\la)\\ %[-1ex]
{\rm F}_-(\om|\la)\end{matrix}\bigg),\qquad
{\rm\mathbf A}\cdot{\rm\mathbf B} \equiv A_+B_++A_-B_-.
\]
The fermionic Fock-vacuum $|\mu\rangle\!\rangle$
is defined by the conditions
\begin{equation}\label{fermisea}
\begin{aligned}
{\bfc}^{\phantom{\dagger}}(\om)\,|\mu\rangle\!\rangle\;=\;0
~~~{\rm for}~~ \om>-\mu,\\
{\bfc}^{{\dagger}}(\om)\,|\mu\rangle\!\rangle\;=\;0~~~{\rm for}~~ \om<-\mu.
\end{aligned}
\end{equation}
The Hilbert space $\CH$ of the theory is then defined as the
completion of the dense subspace spanned by vectors of the
form
\[
{\mathsf c}[{\mathbf f_n}]\cdots {\mathsf c}[{\mathbf f_1}]\,
{\mathsf c}^{\dagger}[{\mathbf g}_m]
\cdots{\mathsf c}^{\dagger}[{\mathbf g}_1]\,
|\,\mu\,\rangle\!\rangle\,,
\]
where
\[
{\mathsf c}[{\mathbf f}]\,=\,\int_{\BR} d\omega\;{\mathbf f}(\omega)\cdot
{\mathbf c}(\omega),
\quad
{\mathsf c}^{\dagger}[{\mathbf g}]\,=\,\int_{\BR} d\omega\;
{\mathbf g}(\omega)\cdot
{\mathbf c}^{\dagger}(\omega),
\]
with ${\mathbf f}(\omega)$, ${\mathbf g}(\omega)$ smooth and rapidly
decaying at $\pm\infty$.
The resulting Hilbert space decomposes into
sectors with a definite fermion number:
\begin{equation}
\CH\;=\;\bigoplus_{n\in\BZ}\,\CH_{n}\, .
\end{equation}
The $\CH_{n}$ are eigenspaces of the fermion number operator
\begin{equation}
{\mathsf N}\,\df\,\int_{-\mu}^{\infty}d\om \,
\bfc^{\dagger}(\om)\cdot{\bfc}^{\phantom{\dagger}}(\om)+
\int_{-\infty}^{-\mu}d\om \,
{\bfc}^{\phantom{\dagger}}(\om)\cdot\bfc^{\dagger}(\om)\,.
\end{equation}
The second-quantized fermionic field operators are then defined
as
\begin{equation}
\begin{aligned}
\Psi(\la,t)& \;=\; % \sum_{p=\pm}
\int d\om\;\,e^{+i\om t}\;{\rm\mathbf F}(\om|\la)\cdot
{\bfc}^{\phantom{\dagger}}(\om)
% +{\rm\mathbf F}_{\om}^-(\la)\,{\bfc}^{\phantom{\dagger}}_{-}(\om)~\big)
~,\\
\Psi^{\dagger}(\la,t)& \;=\;%\sum_{p=\pm}
\int d\om\;\,e^{-i\om t}\;{\bfc}^{\dagger}(\om)\cdot{\rm\mathbf F}(\om|\la)
%+{\rm\mathbf F}_-(\om|\la)\,{\bfc}^{\dagger}_{-}(\om)~\big)
~.
\end{aligned}
\end{equation}
The dynamics of the theory is generated by the Hamiltonian
\begin{equation}
\SH\,=\,\int d\la \;\Psi^\dagger\left(-\frac{\partial^2}{\partial \la^2}-\frac{1}{4}
\la^2\right)\Psi\,.
\end{equation}

As usual in fermionic field theories one may construct observables as bilinear
expressions in the fermionic fields. One may e.g. consider the collective
field
\begin{equation}\label{collfield}
\pa_\la
\chi(\la,t)=\Psi^\dagger(\la,t)\Psi(\la,t)-\langle\!\langle\mu|\Psi^\dagger(\la,t)\Psi(\la,t)|\mu\rangle\!\rangle\,.
\end{equation}
The dynamics generated by the Hamiltonian $\SH$ becomes rather complicated in terms of the
field $\chi$. Consideration of observables like \rf{collfield}
will therefore only be useful in certain limiting cases.

\subsection{In- and Out-fields}

A crucial feature of the inverted harmonic oscillator potential
is that the asymptotic behavior of the wave-function $\psi(\la,t)$
for late/early times can be represented in a very simple
way:
\begin{equation} \psi(\lambda, t)
\quad {}_{\widetilde{t \rightarrow \pm\infty}}
\quad (2 \pi)^{-\frac{1}{2}} e^{\frac{i}{4} \lambda^2} \,
e^{\mp\frac{t}{2}}\,
\chi_{\pm}(u_\pm),
%\qquad\begin{aligned}& s\equiv \sgn(\la)\\
%& x\equiv\ln|\la|.\end{aligned}
\label{asympt}
\end{equation}
where $u_{\pm}\equiv \la e^{\mp t}$.
A proof of this claim is
given in appendix \ref{AppIO}.This means that asymptotically
for $t\ra\pm \infty$ the time evolution becomes represented
by scale transformations. In terms of the coordinate
$x=\ln|\la|$ one finds free relativistic motion.

The  asymptotic wave-functions $\chi_{\pm}(u_\pm)$ can be
calculated from the wave function $\psi(\la)\equiv \psi(\la,0)$
by means of the integral transformations
\begin{equation}\label{inttrsf}
\chi_{\pm}^{}(u_\pm^{})\,\equiv\,(\SM_\pm\psi)(u_\pm^{})\,\equiv\,
\int_{\BR}d\la\;K_{\pm}^{}(u_\pm^{}|\la)\, \psi(\la),
\end{equation}
with kernels $K_{\pm}(u_\pm|\la)$ given by the explicit formulae
\begin{equation}\label{intkern}
K_{+}(u_+|\la)\,=\,
e^{-i\frac{\pi}{4}}e^{\frac{i}{2}u^2_+-i\la u_++\frac{i}{4}\la^2},
\quad
K_{-}(u_-|\la)\,=\,K_{+}^*(u_-|\la).
\end{equation}

These observations lead to a natural
definition of the in- and out-fields. Let us
consider the asymptotics
for $t\ra\pm\infty$ of the fermionic operators
\begin{equation}\label{timefermi}
\Psi^{\dagger}[\,\psi\,|\,t\,)\,\equiv\,
\int d\lambda\; \psi(\la)\,\Psi^{\dagger}(\la,t)=
\int d\lambda\; \psi(\la,t)\,\Psi^{\dagger}(\la).
\end{equation}
It is then natural to define the in- and out fields
$\Psi^{\dagger}_{\pm}(u)$ by the asymptotic relation
\begin{equation}\label{Psiasy}
\Psi^{\dagger}[\,\psi\,|\,t\,)\;\underset{t\ra\pm\infty}{\sim}
\;\Psi^{\dagger}_{\pm}[\,\chi\,|\,t\,)\equiv
\int \frac{du}{2\pi}\; \chi_{\pm}(u e^{\mp t})\,
\Psi^{\dagger}_{\pm}(u),
\end{equation}
where $\chi_{\pm}$ is defined in \rf{asympt}.
It is shown in the   appendix \ref{AppIO} that
$\Psi^{\dagger}[\,\psi\,|\,t\,)$ indeed has asymptotics of the
required form \rf{Psiasy}, with $\Psi^{\dagger}_{\pm}(u_\pm^{})$
being related to $\Psi^{\dagger}(\la)$ by the integral transformations
\begin{equation}
\Psi^{\dagger}_{\pm}(u_\pm^{})=\int_\BR d\la\;
K^*_{\pm}(u_\pm^{}|\la)\Psi^{\dagger}(\la)\,.
\end{equation}
The transformation between in-
and the out-field becomes particularly simple,
\begin{equation}\label{PsiFoutf}
\Psi_{+}(u_+^{})\,=\,\frac{1}{\sqrt{2\pi}}
\int_\BR du_- \;e^{-iu_+u_-}\,
\Psi_{-}(u_-^{})\,.
\end{equation}

It is useful to translate this into the energy representation.
The observation that the transformation \rf{inttrsf}
maps the single-particle
Hamiltonian $\sh$ into the generator for scale transformations
of the coordinates $u_\pm$ makes it easy to find the
expansion into energy
eigenfunctions:
\begin{equation}\label{umodes}
 \Psi^{\dagger}_{\pm}(u_\pm)\;=\;
\int_{\BR} \frac{d\om}{\sqrt{2\pi}}\;
|u_{\pm}|^{\pm i\om-\frac{1}{2}}\,
\big(\Theta(-u_\pm)
\rmd^{\dagger L}_{\pm}(\om)+\Theta(u_\pm)
\rmd^{\dagger R}_{\pm}(\om)\big)
\end{equation}
where $\Theta(u)$ is the usual step function.
The operators $\rmd^{\dagger L}_{\pm}(\om)$ and
$\rmd^{\dagger R}_{\pm}(\om)$ create
fermions which are asymptotically located either
on the right or left
of the potential $V=-\frac{1}{4}\la^2$, respectively.
It will be convenient
to regard $\rmd^{\dagger L}_{\pm}(\om)$ and
$\rmd^{\dagger R}_{\pm}(\om)$ as the two
components of a vector $\bfd^{\dagger}_{\pm}(\om)$.
The relation between the oscillators
$\bfd^{\dagger}_{\pm}(\om)$ and
$\bfc^{\dagger}(\om)$ takes
the form
\begin{equation}\label{oscrel}
\bfd^{\dagger}_{\pm}(\om)=
{\rm \mathbf M}^{\dagger}_\pm(\omega)\cdot\bfc^{\dagger}(\om),
\qquad
\bfd^{\dagger}_{+}(\om)=
{\rm \mathbf R}^{\dagger}(\omega)\cdot\bfd^{\dagger}_{-}(\om)\,,
\end{equation} where the matrices ${\rm \mathbf M}_\pm$
represent the unitary operators ${\mathsf M}_\pm$ defined in
\rf{inttrsf} in the energy representation, and
\begin{equation}\label{Frefl}
{\rm\mathbf R}(\omega)\,\equiv({\rm \mathbf M}_+(\omega))^{2}\,
=\bigg(\begin{matrix}
\rho(\omega) & \theta(\omega) \\ % [-.75ex]
 \theta(\omega) & \rho(\omega)
\end{matrix}
\bigg)\,,
\end{equation}
with matrix elements $\rho(\omega)$, $\theta(\omega)$ having
the following explicit expressions:
\begin{equation}\label{rhoom}
\rho(\omega)=\frac{1}{\sqrt{2\pi}}e^{-\frac{\pi}{2}\omega}
\Ga\big(\fr{1}{2}-i\omega\big),\qquad
\theta(\omega)=-ie^{\pi\omega}\rho(\omega).
\end{equation}
Note that ${\rm\mathbf R}$
is the matrix which represents the reflection of a single fermion in the
potential $V=-\frac{1}{4}\la^2$.
The definition of in- and out-fields leads straightforwardly
to the definition of the unitary S-operator, which may be represented
as
\begin{equation}
{\mathsf S}\,=\,\exp\left(-\int_{\BR}d\om \;\bfd^{\dagger}_-(\om)\cdot
\log{\rm\mathbf R}(\omega)\cdot \bfd^{\phantom{\dagger}}_-(\om)\right).
\end{equation}

\noindent{\bf Remark: } The formalism presented above is clearly
closely related to the light cone formalism of \cite{AKK}. What
seems to be new is our proof (appendix A) of the equivalence
between this formalism and the conventional definition of in- and
out fields in terms of time-asymptotics. This explains the
observation in \cite{AKK} that the Fourier transformation
\rf{PsiFoutf} correctly describes the scattering of free fermions
in the inverted harmonic oscillator potential.

\subsection{Scattering of the bosonic excitations}

However, we are also interested in the
bosonic fields ${\rm S}_{\pm}(x)$ which are defined by bosonizing the
in- and out fields
$\Psi^{\dagger}_{\pm}$ as
\begin{equation}\begin{aligned}
\label{boson} u_\pm\pa_\pm{\rm S}_{\pm}^{s}(u_\pm) &\,= \,
\int_{\BR} \frac{d\om}{\sqrt{2\pi}}\; |u_{\pm}|^{\pm i\om}\,
\big(\Theta(-u_\pm) \rma^{\rm L}_{\pm}(\om)+\Theta(u_\pm)
\rma^{\rm R}_{\pm}(\om)\big)
\,,\\
\rma_{\pm}^{s}(\om)& \,= \,\int_\BR d\om'\;
\rmd_{\pm}^{s}{}^{\dagger}(\om')\,
\rmd_{\pm}^{s}(\om+\om')\,,\qquad
s\in\{+,-\}\,\widehat{=}\,\{{\rm R},{\rm L}\}
\end{aligned}
\end{equation}
where $\pa_\pm\equiv\frac{\pa}{\pa u_\pm}$.
The operators $\rma_{\pm}^{s}(\om)$
satisfy the following commutation relations,
\begin{equation}\label{CCR}
[\,\rma_{\pm}^{s}(\om)~,~\rma_{\pm}^{s'}(\om')\,]\,=
\,\,\om\,\de^{s,s'}
\,\de(\om+\om').
\end{equation}
With the help of the oscillators $\rma_{\iota}^{s}(\om)$, $\iota\in\{+,-\}$
we may construct the
states
\begin{equation}\label{bos-states}
\rma_{\iota}^{\rm s_n}[f_n] \cdots\rma_{\iota}^{\rm s_1}[f_1]\,
\rma_{\iota}^{\rm t_n}[g_n] \cdots\rma_{\iota}^{\rm t_1}[g_1]\,
|\,\mu\,\rangle\!\rangle\,,\qquad \rma_{\iota}^{\rm s}[h]
\equiv\int d\om\,h(\om)\rma_{\iota}^{s}(\om)
\end{equation}
which generate subspaces $\CH_{\0,\0}^{\pm}$ of $\CH$.
It is important to note, however, that the vectors \rf{bos-states} do not
even exhaust the subspace $\CH_\0\subset\CH$ of fermion number zero.
The operator
\begin{equation}\begin{aligned}
{\mathsf K}_\pm\,\df\,\int_{-\mu}^{\infty}d\om \,\big(&
\rmd_\pm^{\rm\sst R}{}^{\dagger}(\om)
\rmd_\pm^{\rm\sst R}(\om)-\rmd_\pm^{\rm\sst L}{}^{\dagger}(\om)
\rmd_\pm^{\rm\sst L}(\om)\big)\\[-.5ex] & +
\int_{-\infty}^{-\mu}d\om \,\big(
\rmd_\pm^{\rm\sst R}{}(\om)
\rmd_\pm^{\rm\sst R}{}^{\dagger}(\om)
-\rmd_\pm^{\rm\sst L}{}(\om)
\rmd_\pm^{\rm\sst L}{}^{\dagger}(\om)\big)\,.
\end{aligned}
\end{equation}
measures the difference between the numbers of fermions which
asymptotically end up to the right and left sides of the potential
respectively.
This means in particular
that $\CH_\0$ decomposes into ``k-sectors'' \cite{DRSVW}
$\CH_{\0,k}$ as follows:
\begin{equation}\label{k-sectors}
\CH_\0\,=\,\bigoplus_{k\in\BZ}\,\CH_{\0,k}^{\pm}\,.
\end{equation}
In order to generate all of $\CH_\0$ from $|\,\mu\,\rangle\!\rangle$
we also need to consider
operators like
\begin{equation}\label{k-change}
\begin{aligned}
{\rm B}_{\pm}(\om) \,= \,\int_\BR d\om'\;\big(&
K^{\rm\sst LR}(\om|\om')\,\rmd_{\pm}^{\rm\sst L}{}^{\dagger}(\om')\,
\rmd_{\pm}^{\rm\sst R}(\om+\om')\\
+& K^{\rm\sst RL}(\om|\om')\,\rmd_{\pm}^{\rm\sst R}{}^{\dagger}(\om')\,
\rmd_{\pm}^{\rm\sst L}(\om+\om')\big)\,.
\end{aligned}
\end{equation}
The S-operator does not map the sector $\CH_{\0,k}^{\rm in}$
to $\CH_{\0,k}^{\rm out}$. In order to see this, let us notice that
inserting \rf{oscrel} into \rf{boson} yields a relation of the
form
\begin{equation}\label{aoutsplit}
\rma_{\rm out}^{s}(\omega)\,=\,[\,\rma_{\rm out}^{s}(\omega)\,]_{\0,\rm in}^{}+
[\,\rma_{\rm out}^{s}(\omega)\,]_{\0,\rm in}^{\perp}
%{\rm b}_{\rm out}^{s}(\omega)
\end{equation}
where $[\rma_{\rm out}^{s}(\omega)]_{\0,\rm in}^{}$ preserves
the sectors $\CH_{\0,k}^{\rm in}$, whereas
$[\,\rma_{\rm out}^{s}(\omega)\,]_{\0,\rm in}^{\perp}$ is of the form \rf{k-change}. The
term $[\rma_{\rm out}^{s}(\omega)]_{\0,\rm in}^{}$
is dominant for $|\omega|\ll \mu$, in which case
either pure reflection or pure transmission dominate the
fermionic scattering of particle-hole pairs.

The perturbative (in $\mu^{-1}$) part of the bosonic S-matrix
is encoded in
\begin{align}
 & R^{\sst(m\mapsto n)}(\underline{\omega}_1,\dots,
\underline{\omega}_n|
\underline{\omega}'_1,\dots,\underline{\omega}'_m)\,=\,\nonumber \\
& \qquad\qquad=\, \langle\!\langle \,\mu\,|\, \rma_{\rm\sst
out}^{s_1}({\omega}_1) \dots \rma_{\rm\sst
out}^{s_n}({\omega}_n)\, \rma_{\rm\sst
in}^{s_1'}(-{\omega}_1')\dots \rma_{\rm\sst
in}^{s_m'}(-{\omega}_m')\, |\,\mu\,\rangle\!\rangle\,,
\end{align}
where we have abbreviated
$\underline{\omega}\equiv(\om,s)$. These matrix elements
are unambiguously defined by \rf{oscexp}, \rf{CCR} together
with $\rma_{\rm out}^{s}(\omega)|\,\mu\,\rangle\!\rangle=0$ for
$\omega>0$. A diagrammatic formalism for the explicit evaluation of the
S-matrix elements has been developed in \cite{MPR}.
It will be useful for
us to observe that the diagonal\footnote{w.r.t. the
decomposition into k-sectors \rf{k-sectors}}
part of the
scattering of the bosonic
excitations can alternatively be encoded in the
following operator relations:
\begin{equation}\boxed{
\begin{aligned}\label{oscexp}
& [\,\rma_{\rm out}^{s}(\omega)\,]_{\0,\rm in}^{}\,
=\\
&=\sum_{s'=\rm\sst L,R}\, \sum_{n=1}^{\infty}
\int\limits_{-\infty}^{\infty}
\frac{d\om_1}{\om_1}\int\limits_{\om_1}^{\infty}\frac{d\om_2}{\om_2}
\cdots
\int\limits_{\om_{n-1}}^{\infty}\frac{d\om_n}{\om_n}\\
& \hspace{4cm} \ti \, R_{(n)}^{ss'}(\,\om\,|\,\om_1,\dots,\om_n\,)\,
\rma_{\rm in}^{s'}(\omega_1)\cdots
\rma_{\rm in}^{s'}(\omega_n)\,. % \nonumber
\end{aligned}}
\end{equation}
The proof of formula \rf{oscexp} together with the explicit
expressions for the coefficient functions $R_{(n)}^{ss'}$ are
given in appendix \ref{inoutbose}. The amplitude $R^{m\mapsto 1}$
that one can read off directly from \rf{oscexp} is very similar to
the corresponding result of \cite{MO} for type 0A two-dimensional
string theory.

It is also shown in appendix \ref{inoutbose} that the leading
asymptotics of this relation for $\mu\ra\infty$, $\om\ll\mu$ is
given by
\begin{equation}
\begin{aligned}\label{tree-oscexp}
& [\,\rma_{\rm out}^{s}(\omega)\,]_{\0,\rm in}^{}\, =\,
\sum_{n=1}^{\infty} \mu^{-i\om+1-n} \int\limits_{-\infty}^{\infty}
{d\om_1}\int\limits_{\om_1}^{\infty}{d\om_2} \cdots
\int\limits_{\om_{n-1}}^{\infty}{d\om_n}\\
& \hspace{4cm} \ti \, \frac{\Ga(1+i\om)}{\Ga(2-n+i\om)}
\de(\om-{\textstyle\sum_{r=1}^n\om_r})\,
\rma_{\rm in}^{s}(\omega_1)\cdots
\rma_{\rm in}^{s}(\omega_n)\,. % \nonumber
\end{aligned}
\end{equation}
This is a trivial generalization of the formula derived in
\cite{Po2,MP}.

% It will finally be useful for us to observe that the completeness
% of the states $|\varpi\rangle\!\rangle_{\pm}^{}$ implies
% the existence of a relation of the form
% \begin{align}\label{out->in}
% & \langle\!\langle \,\mu\,|\,\rma_{\rm\sst out}({\omega}_1)\dots
% \rma_{\rm\sst out}({\omega}_n)\,=\, \\
% &\,\qquad = \;\sum_{m=1}^{\infty}\;\int_{\BR_+}
% d{\omega}'_1 \dots d{\omega}'_m
%  \; R^{\sst(m\mapsto n)}_{(\0)}
% ({\omega}_1,\dots,{\omega}_n|
% {\omega}'_1,\dots,{\omega}'_m)\nonumber\\
% & \hspace{7cm}
%\times\langle\!\langle \,\mu\,|\,\rma_{\rm\sst in}
% ({\omega}_1')\dots
% \rma_{\rm\sst in}({\omega}_m'),\nonumber
% \end{align}

\subsection{Duality conjecture}

From now on let us restrict attention to the excitations supported
on the right of the maximum of the inverted harmonic oscillator
potential. Ignoring the other side will be a good approximation as
long as all energies are well below the top of the potential. The
conjectured duality between $c=1$ string theory and free fermionic
field theory can be formulated most simply in terms of rescaled
bosonic oscillators
\begin{equation}
{\rm b}_{\pm}(\om)\,\equiv\,e^{\pm i\de(\om)}\rma_\pm^{\rm\sst R}
(\om),
\end{equation}
provided the phase
$\de$ is chosen as
\begin{equation}
e^{i\delta(\om)}\,\equiv\,
\frac{\Ga({+}i\omega)}{\Ga(-i\omega)}\,.
\end{equation}
One manifestation of the conjectured duality between the $c=1$
string theory and the free fermionic field theory may then be
formulated as the validity of
\begin{align}\label{cl-duality}
 \langle\!\langle \,\mu\,|\,
\rmb_{\rm\sst out}({\omega}_1) \dots
\rmb_{\rm\sst out}({\omega}_n)\,&
\rmb_{\rm\sst in}(-{\omega}_1')\dots
\rmb_{\rm\sst in}(-{\omega}_m')\,
|\,\mu\,\rangle\!\rangle\,\asymp_{g_s}\,\\
& \,\asymp_{g_s}\,
\big\langle\,T_{\rm\sst out}({\omega}_1)\dots
T_{\rm\sst out}({\omega}_n)\,
T_{\rm\sst in}({\omega}_1')\dots
T_{\rm\sst in}({\omega}_m')\,
\big\rangle_{c=1}^{}\,,\nonumber
\end{align}
where $\asymp_{g_s}$ means equality of asymptotic expansions in
$g_s=\mu^{-1}$. Note that the matrix elements on the left of
\rf{cl-duality} by  themselves do not define a unitary S-matrix,
but the deviation from unitarity is nonperturbative ($\propto
e^{-\frac{1}{g_s}}$),  as follows from \rf{rhoom}.

\section{D0-branes versus fermions - leading order}

Given the duality between the $c=1$ noncritical string theory and
the free fermionic field theory it is natural to ask how to
interpret the fermionic fields within the $c=1$ noncritical string
theory. A proposal for how to answer this question emerged from
\cite{McGV,KMS}: The excitations created by the fermionic fields
can be interpreted as the unstable D0-branes of the $c=1$
noncritical string theory. In the following section we will review
the existing evidence for this identification.

\subsection{D0-branes in type 0B $c=1$ string theory}

The $c=1$ noncritical string theory contains unstable
D0-branes \cite{ZZ}.
These D0-branes are localized in the
strong coupling region $\phi=\infty$. In order to describe their
decay one may consider the boundary interaction
\begin{equation}\label{bdint}
S_{\rm int}\,=\,\kappa\int_{\pa\Sigma}d\tau\; \cosh X_0\,.
\end{equation}
A construction for the corresponding boundary states was
first proposed in \cite{sen}.
These boundary states have to
be tensored with the boundary states for Liouville theory
which describe the D0-branes
\cite{ZZ}.
In this way one arrives at the following
result for the leading order closed string
emission amplitudes:
\begin{equation}\label{worldsheet}
%\begin{aligned}
\langle\,T_{\rm\sst out}(\omega)\,|\,
B_\kappa\,\rangle_{\rm\sst HH}
\,=\,
e^{i\delta(\omega)}
e^{-i\omega\log\sin\pi\kappa}\mu^{-i\frac{\om}{2}},
%\end{aligned}
\end{equation}
The notation $|\,B_\kappa\,\rangle_{\rm\sst HH}$ reminds of the
fact \cite{LLM} that the definition of the boundary state associated to
the boundary interaction \rf{bdint} depends on a choice of
integration contour, $|\,B_\kappa\,\rangle_{\rm\sst HH}$ being
the boundary state associated to the so-called Hartle-Hawking contour
\cite{LLM}.

\subsection{Evidence for the correspondence between D0-branes and fermions}

The authors of \cite{KMS} propose that the state
$|\,\la_\0\,\rangle\!\rangle$ which describes a
fermion with a well-defined initial localization at $\la_\0$
may - at least to leading order in the semiclassical limit -
be represented in the following bosonized form:
\begin{equation}\label{fermiformal}\begin{aligned}
|\,\la_\0\,\rangle\!\rangle\,=\,
:\exp\left(i{\rm S}_{\sst\rm out}
(\la_\0)\right)\!:|\,\mu\,\rr~.
\end{aligned}
\end{equation}
We will later discuss the applicability of the approximation
\begin{equation}
|\,\la_\0\,\rangle\!\rangle\,\simeq\,
\Psi^{\dagger}_{\sst{\rm out}}(\la_\0)\,
|\,\mu\,\rangle\!\rangle
\end{equation}
underlying the  proposal \rf{fermiformal}. Adopting
\rf{fermiformal} as a working hypothesis for the moment,
one seems to find straightforwardly
that
\begin{equation}\label{formaloverlap}
%\begin{aligned}
 \langle\!\langle\,\mu\,|\,\rmb_{\sst\rm out}(\omega)\,
|\,\la_\0\,\rangle\!\rangle
\,=\,
e^{i\delta(\omega)}
e^{-i\omega\log \la_\0}.
%\end{aligned}
\end{equation}
This matches the result of the
worldsheet-computation provided that the initial location $\la_\0$ of the
fermion is related to the parameter $\kappa$ of the unstable ZZ-brane via
\begin{equation}\label{kappa->lam}
\la_\0\,=\,\sqrt{\mu}\sin\pi\kappa.
\end{equation}
The precise match of \rf{formaloverlap} with the
worldsheet-computation for the closed string emission from a
decaying ZZ-brane represents evidence for the identification of
the single fermion state with the ZZ-brane.

\subsubsection{The UV problem}

So far we have been considering the state
$|\,\la_\0\,\rangle\!\rangle$ which corresponds to a definite
D0 brane parameter $\kappa$ via
\rf{kappa->lam}.
However, this state is clearly not normalizable. It was pointed
out in \cite{KMS} that the resulting divergence
of energy expecation values accounts for the corresponding
singular behavior in the expectation values of the
energy emitted from a decaying D-brane as discussed in
\cite{LLM}.
The natural way to resolve this problem is to
average over the initial localization with a
given wave-function $\vf(\la_\0)$,
\begin{equation}
|\,\vf\,\rangle\!\rangle\,\equiv\,
\int d\la_\0 \;\vf(\la_0)\,|\,\la_\0\,\rangle\!\rangle
\end{equation}
Indeed, the norm of the resulting state will be bounded by
the norm of the wave-function $\vf$, making
it obvious that the ultraviolet problem is
resolved.

\subsubsection{The IR problem}

On the other hand one must observe that the overlaps on the left
hand sides of \rf{formaloverlap} are identically zero since
fermion number is conserved in the free fermionic field theory. To
be more explicit, let us note that $\Psi^{\dagger}_{\sst{\rm
out}}(\la_0)\, |\,\mu\,\rangle\!\rangle\in\CH_{\1}$, whereas
$\rma_{\sst\rm out}({\omega})|\,\mu\,\rangle\!\rangle\in\CH_{\0}$.
This implies that the overlaps in \rf{formaloverlap} are indeed
identically zero. There is no contradiction with \rf{fermiformal}
since the bosonization formula \rf{fermiformal} has serious
infrared problems\footnote{In Mandelstam's work one was dealing
with a {\it massive} theory!}.

The aim of the next section is to discuss how to resolve this
puzzle and how to reconcile
the essence of the
proposal of \cite{KMS} with the fermion number conservation
in the free fermionic field theory. More precisely,
we will propose answers to the following two questions:
\begin{itemize}
\item[$\triangleright$] What are reasonable nonvanishing analogs
of the amplitudes \rf{formaloverlap}?
\item[$\triangleright$] What is the proper string-theoretic
interpretation of these amplitudes? Can they be interpreted
in terms of ZZ-brane decay?
\end{itemize}

\section{Fermions vs. solitons}

We are now going to explain how to resolve
the IR problem that was
pointed out at the end of previous section.

\subsection{Solitonic sectors}

To begin with, let us interpret the sectors $\CH_n$ from the
bosonic perspective. To simplify the notation let us temporarily
restrict attention to the in-fields $\Psi(x)\equiv
e^{\frac{x}{2}}\Psi_{-}(e^x)$ and
 $S(x)\equiv S_{-}(e^x)$.

An important point to observe is the fact that the different
sectors can only be distinguished with the help of {\it global}
observables. States $|\vf\rangle\!\rangle_\1^{}$ in $\CH_1$ can be
created from $|\mu\rangle\!\rangle$ via
\begin{equation}
|\vf\rangle\!\rangle_\1^{}\,=\,\Psi^{\dagger}[\vf]\,
|\mu\rangle\!\rangle,\quad
\Psi^{\dagger}[\vf]\,\equiv\,\int_{\BR}dx \,\vf(x)\Psi^{\dagger}(x)\,.
\end{equation}
We shall analyze the physical content of the
states $|\vf\rangle\!\rangle_\1^{}$ from the
bosonic perspective - the observables used to
measure properties of the states $|\vf\rangle\!\rangle^{}_\1$ will, as usual,
be constructed out of the {\it bosonic} oscillators $\rma(\omega)$.
It is not terribly difficult to
show  that
\begin{equation}\label{kink}
%\begin{aligned}
{}\quad \langle\,\pa_x{\rm S}(x)\,\rangle_{\vf}\equiv
{}_\1^{}\langle\!\langle\,\vf\,|\,\pa_x{\rm S}(x)\,|\,
\vf\,\rangle\!\rangle_\1^{}\,
>\,0\quad\forall x\in\BR.
%\\
%{\rm (ii)}& \quad \lim_{\Lambda\ra\infty}
%\int_{-\Lambda}^{+\Lambda}dx \;
%\langle \,\pa_x {\rm S}(x)\,\rangle=1.
%\end{aligned}
\end{equation}
This means that the states $|\vf\rangle\!\rangle_\1$ are solitonic
in the classical sense: They describe kinks of the bosonic field
${\rm S}$.
The states
$|\vf\rangle\!\rangle_\1$ differ in their {\it global}
properties
from any state $|\vf\rangle\!\rangle_\0\in\CH_\0$. The latter satisfy
\begin{equation}
\lim_{\Lambda\ra\infty}
\int_{-\Lambda}^{+\Lambda}dx \;
{}^{}_\0\langle\!\langle\,\vf\,|\,\pa_x{\rm S}(x)\,|\,
\vf\,\rangle\!\rangle_\0^{}\,=0.
\end{equation}
This should be compared to
the expectation value taken in the state
$|\vf_\1\rangle\!\rangle$,
\begin{equation}
\lim_{\Lambda\ra\infty}
\int_{-\Lambda}^{+\Lambda}dx \;
{}^{}_\1\langle\!\langle\,\vf\,|\,\pa_x{\rm S}(x)\,|\,
\vf\,\rangle\!\rangle_\1^{}\ =1.
\end{equation}
The difference between the asymptotic values of ${\rm S}$
measures the number of solitons $\equiv$ fermions.

%sort of coherent state of closed string radiation.

Nevertheless, as long as one uses only {\it local} observables
to measure properties of the states   $|\vf\rangle\!\rangle_n$, $n=0,1$
one will not be able to determine which sector $\CH_n$ a given
state $|\vf\rangle\!\rangle$ belongs to. It is impossible to
measure the soliton charge by using local observables like
$\pa{\rm S}[f]\df
\int_\BR dx \, f(x)\pa_x{\rm S}(x)$ for $f(x)$ nonzero
only in a compact subset of $\BR$. This point can be understood
quite clearly by looking back at \rf{kink}. Imagine
we are measuring
\[
\langle\,\pa{\rm S}[f]\,\rangle_{\vf}\,\equiv\,
{}^{}_\1\langle\!\langle\,\vf\,|\,\pa{\rm
S}[f]\,|\,\vf\,\rangle\!\rangle^{}_\1
\] for
$f>0$ having support in small intervals. If
$|\vf\rangle\!\rangle\in\CH^{}_\1$ we will find a positive result
for whatever interval we have chosen. After having performed
such measurements
for a large number of different intervals one may feel inclined
to say that the probability that the state under
consideration is solitonic is rather high.
Nevertheless one can never be sure that one
will {\it always} find a positive result if one was able to continue the
measurements ad infinitum.

One may therefore regard the sectors $\CH_0$ and $\CH_1$ as
physically equivalent as long as only measurements of {\it local}
observables are concerned. However, mathematically the sectors are
not equivalent at all. This is illustrated most clearly by the
fact that the sector $\CH_1$, as opposed to $\CH_0$, does not
contain a normalizable ground state (state of energy $-\mu$), as
proven in appendix B. This implies that the sectors $\CH_\1$ and
$\CH_0$ are {\it not} unitarily equivalent as representations of
the algebra generated by the $\pa S[f]$. The mathematical
non-equivalence between the sectors becomes physically relevant as
soon as a physical meaning is assigned to {\it global} observables
like the difference between the asymptotic values of the scalar
field.

\subsubsection{Approximate vacua}

The discussion of the previous subsection may be reformulated in
terms of energies as the the statement that we are unable to
distinguish states in $\CH_0$ and $\CH_1$ as long as our detectors
are insensitive to states below a certain minimal energy $\om_{\rm
min}$.

If however, as is usually the case, one is interested in measuring
local observables only, it is perfectly sufficient to have states
within $\CH_1$ which resemble the ground state
$|\,\mu\,\rr\in\CH_0$ to any given accuracy. A simple example for
such states are the vectors
\begin{equation}\label{yketdef}\begin{aligned}
|\,y\,\rangle\!\rangle_\1^{}\,&\equiv\,(2y)^{\frac{1}{2}}\,
e^{-\mu y}\,
\Psi^{\dagger}(-iy)\,|\,\mu\,\rangle\!\rangle\,\\
&\equiv\,(2y)^{\frac{1}{2}}\int_{-\mu}^{\infty}d\om \;
e^{-(\om+\mu)y}\,{\rm d}^{\dagger}(\om)\,|\,\mu\,\rr,
\end{aligned}\qquad y >0, \end{equation}
for large positive values of $y$.
It is easy to show that
\begin{equation}\label{apprvac}
\begin{aligned}
{\rm (i)}& \quad
{ {}_\1^{}\langle\!\langle \,y\,|\,\textsf{H}\,
|\,y \,\rangle\!\rangle_\1^{} }\,=\,(2y)^{-1},\\
{\rm (ii)}& \quad {{}_\1^{}\langle\!\langle \,y\,|\,\pa_x{\rm S}(x)
\,|\,y\,\rangle\!\rangle_\1^{}}\,=\,
\frac{2y}{x^2+y^2}
%\,=\,
%\frac{2}{\Im z}\Bigg[1+\Big(\frac{x-\Re z}{\Im z}\Big)^2\Bigg]^{-1}.
\end{aligned}\end{equation}
For given sensitivity of our detectors
we only need to make $y$ large enough to get states
which resemble the bosonic vacuum $|\,\mu\,\rangle\!\rangle$ as much as
we want. We will call such states approximate vacua.
Equation \rf{apprvac}, (ii) offers an intuitive picture
of these states: The profile of the expectation value of
$\pa_x{\rm S}(x)$ becomes arbitrarily flat.

One may then consider states like
$\rma(-\omega)|\,y\,\rangle\!\rangle^{}_\1$. By generalizing
the previous discussion slightly one may convince oneself that
such a state resembles a single boson state. Overlaps like
\[
{}^{}_\1\langle\!\langle\,y\,|\,\rma(\omega)\,|\,\vf\,\rangle\!\rangle^{}_\1
\]
will be nonvanishing and can be interpreted as the amplitude for
transition of a single fermion state into a state which
resembles a state with a single boson created from the vacuum.
For large $y$ one finds
\begin{equation}\label{yPsioverlap}
{}^{}_\1\langle\!\langle\,y\,|\,\rma(\omega)\,\Psi^{\dagger}(x)\,
|\,\mu\,\rangle\!\rangle^{}_\1
\;{\simeq}_y\;(2y)^{-\frac{1}{2}}e^{-i(\omega-\mu) t},
\end{equation}
where the notation $\simeq_{y}$ means equality up to an error controlled
by $y^{-1}$.

\subsubsection{Partial bosonization}\label{Spartbose}

% We are now going to show that the states
% $|\,\vf\,\rangle\!\rangle^{}_\1\in\CH_\1$
% {\bf do} have a description as a coherent state of
% bosonic excitations created from an approximate vacuum in the
% $\CH_\1$-sector.

As a convenient formal device to deal with the infrared problem
of the usual boson\-ization formula we propose to
replace it by the following
``partial bosonization'' formula
\begin{equation}\label{partbose}
\Psi^{\dagger}(x)\,=\,e^{iT_<(x|y)}\,\Psi^{\dagger}(y)\,e^{iT_>(x|y)},
\end{equation}
where
\begin{equation}
T_{>}(x|y)\,\equiv\,\int_0^{\infty}
\frac{d\omega}{\omega}\,(e^{i\omega x}-e^{\omega y})
\rma(\omega)\,,
\end{equation}
and similarly for $T_<(x|y)$. The
states
\[
\Psi^{\dagger}(x)|\,\mu\,\rangle\!\rangle\,\propto\,
e^{iT_<(x|y)}|\,y\,\rangle\!\rangle^{}_\1\,
\]
are then naturally interpreted as coherent states of bosons
created from an approximate vacuum within $\CH_\1$.

It is quite clear that $y$ plays the role of an IR cutoff in
formula \rf{partbose}. It may be puzzling that one can not remove
the cut-off $y$ from transition amplitudes such as
\rf{yPsioverlap}. The small prefactor $(2y)^{-\frac{1}{2}}$ on the
right hand side of \rf{yPsioverlap} implies that the probability
for finding only a single boson in the state $\Psi^{\dagger}(x)\,
|\,\mu\,\rangle\!\rangle$ would vanish if we would try to remove
the IR cutoff $y$. This is physically perfectly appropriate: The
state $|\,y\,\rr$ represents the unobservable "cloud" of radiation
with energy too small to be detected. An increase of $y$ means our
detector was replaced by a better one which can detect quanta of
lower energy. The probability to observe {\it only one} closed
string will be lowered.

%Nonvanishing amplitudes like
%\rf{yPsioverlap} can only be defined if the definition
%of the state ${}^{}_\1\langle\!\langle\,y\,|\,\rma(\omega)$
%includes part of the ``infrared cloud'' that is contained
%in the state $\Psi^{\dagger}(x)|\,\mu\,\rangle\!\rangle$.

\subsection{String-theoretical interpretation}\label{interpret}

Let us now apply the discussion in the previous subsections to the
case of the 2d string theory. It is natural to consider the
following transition amplitude as the proper representative of the
amplitude for emission of a single tachyon from a decaying ZZ
brane within the fermionic field theory:
\begin{equation}\label{overlap}
%\begin{aligned}
 \langle\!\langle\,y\,|\,\rmb_{\sst\rm out}(\omega)\,
|\,\la_\0\,\rangle\!\rangle
\,=\,
e^{i\delta(\omega)}
e^{-i\omega\log\la_\0}\,\lala\,y\,|\,\la_\0\,\rr,
\end{equation}
where
\begin{equation}\label{yla}
\lala\,y\,|\,\la_\0\,\rr\,=\,\sqrt{2y}
\frac{e^{i\mu\log\la_\0}}{y+i\log\la_\0}\,
\simeq_y\,(2y)^{-\frac{1}{2}}\,e^{i\mu\log\la_\0}.
\end{equation}
When trying to interpret this result as the amplitude for emission of a closed
string from a decaying D-brane one may be bothered by the
additional factor $\lala\,y\,|\,\la_\0\,\rr$ in \rf{overlap}.
However, we shall
imagine performing a
gedankenexperiment
in which the radiation from a decaying ZZ-brane is measured with
the help of a tachyon detector installed in the weak coupling
region $\phi\ra-\infty$ of our two-dimensional space time.
More generally
one may consider amplitudes like
$\langle\!\langle\,y\,|\,\CO\,
|\,\la_\0\,\rangle\!\rangle$ for an arbitrary local bosonic observable
$\CO$. {\it Any} such
amplitude will be proportional to $\lala\,y\,|\,\la_\0\,\rr$,
leading us to the conclusion that this overall factor is not
physically relevant for the gedankenexperiment we
are considering.

To exhibit the content of the formulae above from the perspective of
2d string theory let us assume that our tachyon detector measures the expectation value of the
tachyon field as a function of the time $t$,
\begin{equation}
\langle\,\pa_t T_{\sst\rm out}(t)\,\rangle_{\chi}^{}\equiv
{}_\1^{}\langle\!\langle\chi_{\rm\sst out}^{}|\,
\pa_t T_{\sst\rm out}(t)\,|
\chi_{\rm\sst out}^{}\rangle\!\rangle_\1^{}.
\end{equation}
The expectation value is evaluated in a state
$|\chi_{\rm\sst out}^{}\rangle\!\rangle_\1^{}$, which is defined as
\begin{equation}
|\chi_{\rm\sst out}^{}\rangle\!\rangle_\1^{}\,=\,\int_{\BR}dx \,
\chi_{\rm\sst out}^{}(x)
\Psi^{\dagger}_{\rm\sst out}(x)\,|\,\mu\,\rangle\!\rangle.
\end{equation}
Let us assume for simplicity that the wave-function $\chi_{\rm\sst
out}^{}$ has Gaussian decay away from a narrow interval. The
expectation value $\langle\,\pa_t T_{\sst\rm
out}(t)\,\rangle_{\chi}^{}$ can be calculated from the expectation
value of $\pa_t S_{\sst\rm out}^{\rm\sst R}(t)$ and by taking into
account the so-called leg-pole transformation,
\begin{equation}\label{leg-pole}
\langle\,\pa_t T_{\sst\rm out}(t)\,\rangle_{\la_\0}^{}=
\int\limits_{-\infty}^{\infty}\!\!dx
\; K(t-x)\,
\langle\!\langle\chi_{\rm\sst out}^{}|\pa_x
S_{\sst\rm out}^{\rm\sst R}(x)
|\chi_{\rm\sst out}^{}\rangle\!\rangle,
\end{equation}
where $K(x)$ is defined by
\[
K(x)=\int_\BR d\om\;e^{i\om x} \, e^{i\delta(\om)}=-\frac{z}{2}J_1(z),
\qquad z\equiv 2 e^{-\frac{y}{2}}.
\]
The expectation value on the right hand side of \rf{leg-pole}
is sharply peaked. Following the discussion in \cite{NP}
one may then conclude that
the resulting profile for
$\langle\pa_t T_{\sst\rm out}(t)\rangle_{\chi}^{}$ will first exhibit
an exponential growth, reach a maximum, and then decay to zero
faster than exponentially.

On the basis of these observations it seems natural to
propose the following physical interpretation in terms
of 2d string theory.
The closed string observer in the weak
coupling region will conclude that he/she has observed the
radiation from the decay of a ZZ-brane.
After some time,
there will be no detectable radiation any more. Most of the
energy of the ZZ-brane went into radiation, the missing bit
not being detectable. Although fermion number conservation
implies  that there is a low energy ``remnant'' hidden
behind the Liouville wall, there will always be a time
after which the existence of a ZZ-brane becomes unobservable.
In this sense the remnant is unphysical, not being distinguishable from
the true vacuum by any local measurement.

There is yet another point of view that one may adopt.
Given that the D-branes are sources for closed strings
one may interpret the
presence of D-branes as a deformation of the original
$c=1$ closed string background.
The fact that the fermion number distinguishes
superselection sectors then translates into the statement
that backgrounds containing D-branes are not small deformations
of the original $c=1$-background but rather distinguished
from it by boundary conditions related to the asymptotic
values of certain fields.
It seems interesting to note that - in contrast to previous
appearances of topological charges in string theory - here we
find that the topological charges are given by asymptotic
boundary conditions in time rather than space.

% We have to observe, however, that these boundary conditions
% are not easily accessible to our closed string observer
% at $\phi\ra -\infty$. Even if the measurement of the closed
% string radiation could be performed for infinitely long time,
% it would still be necessary to ``undo''
% the leg-pole transformation (cf. \rf{leg-pole})
% in order to get a simple description of the topological
% charge in terms of expecation values of the fields
% ${\rm S}_{\rm\sst out,R/L}$.

\section{Manifestation of open-closed duality}

\subsection{General features of the worldsheet description}\label{WS}

We now want to propose a hopefully suggestive formal line
of arguments leading to a proposal which was made in many
discussions (see e.g. the discussion in \cite{GR} and references
therein) of possible world-sheet explanations for
open-closed dualities: The insertion of discs into string-worldsheets
is equivalent to the insertion of a particular on-shell closed
string vertex operator. Summing over disc insertions amounts
to exponentiating the vertex operator, which describes a shift of the
closed string background. In particular we shall try to
identify some issues connected to this line of thought
on which we shall gain some insight by the subsequent
comparison with the results from the free fermionic field theory.

What sort of amplitudes are we looking for? We want to analyze the
particle production by the time-dependent background that is
furnished by the ``decaying'' D0-brane(s). So very schematically we
are interested in
\begin{equation}\label{amplitude}
\big\langle \,T_{\rm\sst out}({\omega}_1)\cdots
T_{\rm\sst out}({\omega}_n)\, \big\rangle^{}_{\rm  D0,\kappa}\,.
\end{equation}
The notation $\langle\cdots \rangle^{}_{\rm\sst D0,\kappa}$ is
supposed to indicate that the expectation value is not taken in
the usual $c=1$ closed string background, but rather in the
modified background obtained by the insertion of a $D0$-brane with
parameter $\kappa$. The standard world-sheet definition of
amplitudes like \rf{amplitude} takes the following schematic form:
\begin{equation}\label{ampl-exp}\begin{aligned}
\big\langle \,T_{\rm\sst out}({\omega}_1)\cdots &
T_{\rm\sst out}({\omega}_n)\,
\big\rangle^{}_{\rm D0,\kappa}\,\asymp_{g_s}\,\\
& \asymp_{g_s}\,\sum_{h=0}^{\infty}\,\sum_{d=1}^{\infty}\,
g_s^{2h-2+d}\;\big\langle \,T_{\rm\sst out}({\omega}_1)\cdots
T_{\rm\sst out}({\omega}_n)\,
\big\rangle_{c=1.}^{\sst(h,d)}
\end{aligned}\end{equation}
The notation $\asymp_{g_s}$ means equality of asymptotic
expansions in powers of $g_s$. The terms $\langle T_{\rm\sst
out}({\omega}_1)\cdots T_{\rm\sst out}({\omega}_n) \rangle_{\sst
h,d}$ in the expansion \rf{ampl-exp} are associated to Riemann
surfaces $\Sigma_{h,n,d}$ with genus $h$, $n$ punctures and $d$
discs. In general one might try to represent these terms as
integrals over the moduli space $\CM_{h,n,d}$ of Riemann surfaces
$\Sigma_{h,n,d}$,
\begin{equation}\label{modint}\begin{aligned}
\big\langle \,T_{\rm\sst out}({\omega}_1) & \cdots
T_{\rm\sst out}({\omega}_n)\,
\big\rangle^{h,d}_{c=1}=\\
&=\int\limits_{\CM_{h,n,d}}\!\! \Omega_{h,n,d}\; \big\langle \,
v_{\rm\sst out}({\omega}_1)\ot \cdots\ot v_{\rm\sst
out}({\omega}_n)\, \big\rangle^{\rm\sst CFT}_{\Sigma_{h,n,d}(M),}
\end{aligned}\end{equation}
where we have put the ghost contributions into the definition of
the top form $\Omega_{h,n,d}$ and $\langle \cdots\rangle^{\rm\sst
CFT}_{\Sigma_{h,n,d}(M)}$ is a correlation function in the
conformal field theory
\[
{\rm CFT}=({\rm Super-Liouville})\ot (X_0{-\rm CFT}).
\]
The correlation functions $\langle \cdots\rangle^{\rm\sst
CFT}_{\Sigma_{h,n,d}(M)}$ are viewed as machines which for each
point $M\in\CM_{h,n,d}$ transform vectors $v \in\CH_{\rm\sst
CFT}^{\ot n}$ into numbers.

We do not expect unusual problems in the construction of arbitrary
correlation functions $\langle \cdots\rangle^{\rm\sst
CFT}_{\Sigma_{h,n,d}(M)}$ as long as $d=0$. The potentially
troublesome $X_0$-CFT is free in the bulk, which allows us to
define the contribution from the $X_0$-CFT by means of analytic
continuation w.r.t. the energies $\om_k$, $k=1,\dots,n$.
%\footnote{We should point out,
%however, that we are not aware of a reference which analyzes this issue
%in detail.}
In order to construct the amplitudes with disc insertions a
standard approach would be to start from correlation functions
$\langle \cdots\rangle^{\rm\sst CFT}_{\Sigma_{h,n+d,0}(M)}$, from
which one may try to construct $\langle \cdots\rangle^{\rm\sst
CFT}_{\Sigma_{h,n,d}(M)}$ by sewing punctured discs to $d$ of the
$n+d$ punctures. In the case $d=1$ this would lead to a
representation of the following type
\begin{align}\label{discinsert}
\big\langle \, v_{\rm\sst out}({\omega}_1)\ot &  \cdots\ot
v_{\rm\sst out}({\omega}_n)\, \big\rangle^{\rm\sst
CFT}_{\Sigma_{h,n,\1}(M)}
= \\
& =\big\langle \, v_{\rm\sst out}({\omega}_1)\ot \cdots\ot
v_{\rm\sst out}({\omega}_n)\ot
(e^{-\tau(L_0+\bar{L}_0-2)}|B_\kappa\rangle)\,
\big\rangle^{\rm\sst CFT}_{\Sigma_{h,n+1,0}(M),} \nonumber
\end{align}
where $|B_\kappa\rangle$ is the boundary state associated to the
boundary interaction \rf{bdint}. The gluing parameter $\tau\in\BR_+$
represents the deformations of $\Sigma_{h,n,1}$ which change the
radius of the disc. In the general case $d\geq 0$ one will have
$d$ such gluing parameters $\tau_1,\dots,\tau_d$, and an obvious
generalization of formula \rf{discinsert}.

In formula \rf{discinsert} we observe an unusual
source of trouble: The spectrum of $L_0+\bar{L}_0$ is unbounded from
below since the $X_0$-CFT contains eigenstates
with arbitrarily negative eigenvalues. It is therefore
not clear to me how to make sense out of the right hand side of
\rf{discinsert} in the present context.

If, however, a good definition for the right hand side of
\rf{discinsert} is ultimately found, we could proceed with the
integration over moduli space as follows: By using coordinates for
the moduli space $\CM_{h,n,d}$ such as those used in \cite{Ko} one
may realize that
\begin{equation}\label{modfactor}\CM_{h,n,d}\,\simeq\,
 \CM_{h,n+d,0}\ti\BR^d_+,
\end{equation}
where we may think of $\CM_{h,n+d,0}$ as parametrizing the complex
structures on the surface that is obtained from $\Sigma_{h,n,d}$
by gluing punctured discs into the $d$ boundaries of
$\Sigma_{h,n,d}$. The moduli corresponding to the factor $\BR^d_+$
in \rf{modfactor} can be identified with the radii of the discs,
and therfore with the parameters $\tau_1,\dots,\tau_d$ that were
introduced after \rf{discinsert}. This means that we can factor
off the integration over $\BR^d_+$ in \rf{modint} and represent it
explicitly by integrating over $\tau_1,\dots,\tau_d$. As a
symbolic notation for the result of this procedure we shall
propose
\begin{align}\label{disc->punct}
\bigg\langle \,
v_{\rm\sst out}({\omega}_1)\ot \cdots  \ot
v_{\rm\sst out}({\omega}_n)\ot
 & \Big(\frac{1}{L_0+\bar{L}_0-2}\,|B_\kappa\rangle\Big)^{\ot d}
%\cdots \\ & \qquad\cdots\ot
%\Big(\frac{1}{L_0+\bar{L}_0-2}\,|B_\kappa\rangle\Big)\,
\bigg\rangle^{\rm\sst CFT}_{\Sigma_{h,n+d,0}(M).} % \nonumber
\end{align}
The physical interpretation of the insertions of
$({L_0+\bar{L}_0-2})^{-1}|B_\kappa\rangle$ should be clear: They represent
the propagation of closed strings from the brane into the region where
interactions with other closed strings take place. This leads us
to formulate a physically motivated requirement on possible definitions
of \rf{discinsert},\rf{disc->punct}:
They should be such that only on-shell physical
states contribute in \rf{disc->punct}.
We are thereby lead
to the expectation that \rf{modint} can be replaced by an
expression of the following
form
\begin{align}\nonumber
 \big\langle \,T_{\rm\sst out} & ({\omega}_1)  \cdots
T_{\rm\sst out}({\omega}_n)\,\big\rangle^{}_{h,d}=  \\
 & =\!\!\!\int\limits_{\CM_{h,n+d,0}}\!\!\!\! \Omega_{h,n+d,0}\;
\Big\langle \,
v_{\rm\sst out}({\omega}_1)\ot  \cdots\ot
v_{\rm\sst out}({\omega}_n)\ot
% \nonumber \\[-2ex] &\hspace{5cm} \ot
\big(w_{\rm\sst in}(\kappa)\big)^{\ot d}
% \cdots \ot w_{\rm\sst in}(\kappa)\,
\Big\rangle^{\rm\sst CFT}_{\Sigma_{h,n+d,0}(M)}\nonumber\\
& =\,\int\! d{\om}_1'\cdots d{\om}_d' \;\prod_{r=1}^d \;\langle\,
T_{\rm\sst in}  ({\omega}_r')
  \,|\,B_{\kappa}\,\rangle\,\label{modint+} \\[-1ex]
& \hspace{4cm}\times
\big\langle \,T_{\rm\sst out}({\omega}_1)  \cdots
 T_{\rm\sst out}({\omega}_n)\,
              T_{\rm\sst in}({\om}_1')\cdots
T_{\rm\sst in}({\omega}_m')\,
\big\rangle^{\sst (h,0)}_{c=1.}
\nonumber\end{align}
where we have assumed (with some hindsight) that
the state $w_{\rm\sst in}(\kappa)$ may be represented in the form
\begin{equation}
w_{\rm\sst in}(\kappa)\,=\,\int d{\om}\;\langle \,T_{\rm\sst
in}({\omega})\, |\,B_{\kappa}\,\rangle \,v_{\rm\sst in}
({\omega})\,.
\end{equation}
%In \rf{modint+} we have furthermore used the notation
%\[
%\begin{aligned}R^{\sst(m\mapsto n)}_{g}
%({\omega}_1,\dots,{\omega}_n &|
%{\omega}'_1,\dots,{\omega}'_m)=\\
%& =\big\langle \,T_{\rm\sst out}({\omega}_1)  \cdots
%T_{\rm\sst out}({\omega}_n)T_{\rm\sst in}({\om}_1')\cdots
%T_{\rm\sst in}({\omega}_m')\,\big\rangle^{}_{g,0}.
%\end{aligned}
%\]
Equation \rf{modint+} is the sought-for representation of disc
insertions in terms of certain closed string operators.

We believe that the following point deserves some emphasis:
{\em Despite the fact that we do not know
the precise definition for the right hand side of
\rf{discinsert}, we are rather confident that the representation
\rf{modint+} for perturbative closed string emission amplitudes
in terms of a sum over disc insertions should be valid. }

We will soon see that these expectations are nicely supported by
results from the fermionic field theory. This will allow us to
demonstrate that the above ideas about the
world-sheet mechanism behind open-closed duality are realized
in the present context in a rather concrete and well-defined manner.

\subsection{Fermionic definition of amplitudes} \label{Fermidef}
% \subsection{Rolling fermion beyond leading order}

Our aim is to calculate the
amplitudes for emission of closed strings
from a decaying ZZ-brane. Identifying the ZZ-branes with
the fermions in the free fermionic field theory leads one to
consider overlaps of the form
\begin{equation}\label{overlap1}
\langle\!\langle \,y\,|\, \rmb_{\rm\sst out}({\omega}_1)\dots
\rmb_{\rm\sst out}({\omega}_n)\,|\,\vf\,\rangle\!\rangle
\end{equation}
where $|\,\vf\,\rangle\!\rangle$ represents a single fermion
created from the vacuum $|\,\mu\,\rangle\!\rangle$,
\begin{equation}
|\,\vf\,\rangle\!\rangle^{}\,=\, \int d\la
\;\vf(\la)\,|\,\la\,\rangle\!\rangle^{}_\1\, ,\quad
|\,\la\,\rangle\!\rangle^{}_\1 \,\equiv\,\Psi^{\dagger}(\la)
\,|\,\mu\,\rangle\!\rangle.
\end{equation}
To be fully specific let us agree that the pseudo-vacuum
$|\,y\,\rangle\!\rangle$ in \rf{overlap1} is defined as in
\rf{yketdef} by using $\rmd^{\dagger\rm R}_{-}(\om)$ instead of
$\rmd^{\dagger}_+(\om)$.\footnote{It may seem unnatural that we
define $|\,y\,\rangle\!\rangle$ in terms of the in-fermionic
creation operators $\rmd^{\dagger\rm R}_{-}(\om)$ rather than
$\rmd^{\dagger\rm R}_{+}(\om)$. The difference is in many respects
inessential, though. Thanks to the fact that the energy
distribution in \rf{yketdef} is peaked around $\om=-\mu$ we may
approximate the reflection matrix ${\mathbf R}(\om)$ in
\rf{oscrel} by ${\mathbf R}(-\mu)$. This means that replacing
  $\rmd^{\dagger\rm R}_{-}(\om)$ by
$\rmd^{\dagger\rm R}_{+}(\om)$ results in an overall factor that
depends on $\mu$ only, and will therefore be irrelevant for most
questions.
However, this factor
will have to be taken into account when calculating the
asymptotic expansions in $\mu^{-1}$. This will be done in  the following
subsections \ref{CSP} and
\ref{OSP}, where our present choice will turn out to be the most
convenient one.}

We are trying to establish a relation of the form
\begin{equation}\label{comparison}
\langle\!\langle \,y\,|\, \rmb_{\rm\sst out}({\omega}_1)\dots
\rmb_{\rm\sst out}({\omega}_n)\,|\,\vf\,\rangle\!\rangle
\asymp_{g_s} \big\langle \,T_{\rm\sst out}({\omega}_1)\cdots
T_{\rm\sst out}({\omega}_n)\, \big\rangle^{}_{\rm \rm\sst
D0,\kappa}\,.
\end{equation}
This clearly requires choosing a particular wave-function
$\vf\equiv\vf_\kappa$. We know that $|\,\vf\,\rangle\!\rangle$ has
the following equivalent descriptions:
\begin{equation}\label{Dout}\begin{aligned}
|\,\vf\,\rangle\!\rangle\,=\,& \int_{\BR}du_+^{} \,\chi_{\rm\sst
out}^{}(u_+^{})
\Psi^{\dagger}_{\rm\sst out}(u_+^{})\,|\,\mu\,\rangle\!\rangle\\
\,=\, & \int_{\BR}d\la \;\psi(\la)\Psi^{\dagger}(\la)\,
|\,\mu\,\rangle\!\rangle\\
\,=\,&\int_{\BR}du_-^{} \,\chi_{\rm\sst in}^{}(u_-^{})
\Psi^{\dagger}_{\rm\sst in}(u_-^{})\,|\,\mu\,\rangle\!\rangle.
\end{aligned}\end{equation}
Let us furthermore recall that given any one of the wave-functions
$(\chi_{\rm\sst out}^{},\phi,\chi_{\rm\sst in}^{})$,
we can calculate the two others via the integral
transformations \rf{inttrsf}.
These relations describe the dispersion that a wave-packet suffers
in the time-evolution between any finite time and and the
asymptotics $t\ra\pm\infty$. It seems natural to suspect that
the correct choice of $\vf_\kappa$ must correspond to
point-like initial localization, with parameter $\kappa$
being related to the initial position. Still we have two
options to consider: Point-like initial localization at finite
time or point-like initial localization for time $t\ra -\infty$.
These two possibilities are of course inequivalent as the
effects of dispersion will be substantial in general.
We are going to show that only
point-like initial localization for time $t\ra -\infty$
has the chance to yield amplitudes that
can be identified with the world-sheet description. Let us therefore
consider the choice
\begin{equation}\label{chiin}
\chi_{\rm\sst in}^{}(u_-^{})\,\equiv\,
\de(u_-^{}-u_-^\0)\end{equation} which corresponds to choosing
\begin{equation}
|\,\vf\,\rangle\!\rangle\,\equiv\,
|\,u_-^\0\,\rangle\!\rangle\, ,\quad |\,u_-^\0\,\rangle\!\rangle
\equiv
\Psi^{\dagger}_{\sst\rm in}(u_-^\0)\,|\,\mu\,\rangle\!\rangle\,.
\end{equation}
Our next aim will be to show that this choice indeed
leads to a relation of the desired form \rf{comparison}.

\subsection{Closed string picture}\label{CSP}

One possible way to expand the amplitude in powers of $g_s=\mu^{-1}$
proceeds by
using the expansion \rf{oscexp} in order to express the
bosonic operators $\rmb_{\rm\sst out}({\omega})$
in terms of the $\rmb_{\rm\sst in}({\omega})$.
This leads to an expression of the following form:
\begin{align}\label{out->inappr}
& \langle\!\langle \,y\,|\,\rmb_{\rm\sst out}({\omega}_1)\cdots
\rmb_{\rm\sst out}({\omega}_n)\,|\,u_-^\0\,\rangle\!\rangle\,
\asymp_{g_s}\, \\
&\,\quad\asymp_{g_s}\;\sum_{m=1}^{\infty}\frac{1}{m!}\;\,
\int\limits_{\BR_+}
d{\omega}'_1 \;\dots \int\limits_{\BR_+} d{\omega}'_m
 \; R^{\sst(m\mapsto n)}
({\omega}_1,\dots,{\omega}_n|
{\omega}'_1,\dots,{\omega}'_m)\nonumber\\[-2ex]
& \hspace{8cm}
 \times\langle\!\langle \,y\,|\,\rmb_{\rm\sst in}({\omega}_1')\cdots
\rmb_{\rm\sst in}({\omega}_m')\,
|\,u_-^\0\,\rangle\!\rangle\nonumber
\end{align}
We are only claiming equality of asymptotic expansions in
$g_s=\mu^{-1}$ since we have been ignoring the non-perturbative
correction associated to the second term in the decomposition
\rf{aoutsplit}. In the case that we have $\mu\ra\infty$,
$\frac{\om}{\mu}\ll 1$, where $\om\equiv\sum_{r=1}^n\om_r$, we may
approximate $R^{\sst(m\mapsto n)}$ by its asymptotic expansion in
powers of $g_s=\mu^{-1}$. It is important to note that up to terms
of order $y^{-1}$, the S-matrix elements $R^{\sst(m\mapsto n)}$ in
the one-fermion sector are equal to the their counterparts in the
zero fermion sector. The asymptotic expansion of the latter was
identified with the correlation functions of the $c=1$ string
theory in \rf{cl-duality}. This allows us to write
\begin{equation}\label{asymR}
\begin{aligned} R^{\sst(m\mapsto n)} &
({\omega}_1,\dots, {\omega}_n  |
{\omega}'_1,\dots,{\omega}'_m)\asymp_{g_s}\\
\asymp_{g_s}\, & \sum_{h=0}^{\infty}\,g_s^{2h-2+n+m}\,
\big\langle\,T_{\rm\sst out}({\omega}_1)\dots
T_{\rm\sst out}({\omega}_n)\,
T_{\rm\sst in}({\omega}_1')\dots
T_{\rm\sst in}({\omega}_m')\,
\big\rangle^{\sst(h,0)}_{c=1,}
\end{aligned}
\end{equation}
Let us also note the following simple relation
\begin{equation}
\langle\!\langle \,y\,|\,\rmb_{\rm\sst in}
({\omega}_1')\dots
\rmb_{\rm\sst in}({\omega}_m')\,
|\,u_-^\0\,\rangle\!\rangle\,=\,\langle\!\langle \,y\,|
\,u_-^\0\,\rangle\!\rangle\;
\prod_{r=1}^d \,\langle \,T_{\rm\sst in}({\omega}_r')\,
\,|\,B_{\kappa}\,\rangle.
\end{equation}
We thereby arrive at an expansion of the following form
\begin{equation}\boxed{
\begin{aligned}\label{out->inappr2} & \langle\!\langle \,y\,|
\,u_-^\0\,\rangle\!\rangle^{-1}_{} \langle\!\langle
\,y\,|\,\rmb_{\rm\sst out}({\omega}_1)\cdots \rmb_{\rm\sst
out}({\omega}_n)\,|\,u_-^\0\,\rangle\!\rangle\,
\asymp_{g_s}\, \\[1ex]
&\,\asymp_{g_s}\;
\sum_{d=1}^{\infty}\,\frac{1}{d!}\,\sum_{h=0}^{\infty}
g_s^{2h-2+n+d}
\int\limits_{\BR_+}
d{\omega}'_1\; \dots  \int\limits_{\BR_+}  d{\omega}'_d
 \; \,\prod_{r=1}^d \,\langle \,T_{\rm\sst in}({\omega}_r')\,
\,|\,B_{\kappa}\,\rangle\,
 \nonumber \\
& \hspace{4cm}\times\big\langle\,T_{\rm\sst out}({\omega}_1)\dots
T_{\rm\sst out}({\omega}_n)\, T_{\rm\sst in}({\omega}_1')\dots
T_{\rm\sst in}({\omega}_d')\,
\big\rangle^{\sst(h,0)}_{c=1.}\nonumber\end{aligned}}\end{equation}
This is just what we are expecting on the basis of the discussion
in subsection \ref{WS}, cf. in particular with equation \rf{modint+}.\\[1ex]
{\bf Remarks}\\[1ex]
{\bf 1.} The leading asymptotics $\mu\ra\infty$,
$\frac{\om}{\mu}\ll 1$ of the one-point function is in found to be
\begin{equation}\label{onept-as}
\langle\!\langle \,y\,|\,
\rmb_{\rm\sst out}(\omega)
\,|\,F\,\rangle\!\rangle
\, \simeq_y\,(2y)^{-\frac{1}{2}}\,
e^{i\de(\om)}\,e^{-i(\om-\mu)\log u_-^\0}\,\mu^{-i\om}
\end{equation}
Up to inessential factors (cf. our discussion in subsection \ref{interpret})
we find a result that matches the world-sheet computation
provided that we identify the parameters as
\begin{equation}\label{u_in-kappa}
\sin\pi\kappa\,=\,\sqrt{\mu}\,u_-^\0.
\end{equation}
It should be noted that the initial localization $u_-^\0$ has nothing
to do with the turning point of the corresponding classical motion.
The latter may be estimated from the average value of the energy when
we form wave-packets rather than considering point-like localized
``states''. This point will be further elaborated upon 
at the end of the following subsection.\\[1ex]
{\bf 2.} One may/should worry about the convergence of the various
summations/integrations in \rf{out->inappr2}. Let us first note
that the integrations do not pose any problem. The UV convergence
is ensured by the delta-function in the integrand together with
$\lala y|\rmb_{\rm\sst in}({\omega})\sim e^{y\om} \lala y|$ for
$\om<0$. The absence of IR problems can be inferred from formula
\rf{Rdefs}, noting that $Q_{(n)}^{ss'}$ vanishes for $\om_r\ra 0$.
The summation over $h$ will not be convergent but only asymptotic,
but it is interesting to note that the sum over the number of
discs $d$ is probably convergent even {\em after} exchanging the
summations over $h$ and $d$. We have checked this claim explicitly
in the case $n=1$ using the tree approximation
\rf{tree-oscexp} to the closed string S-matrix.\\[1ex]
{\bf 3.} There is another way of writing the expansion
\rf{out->inappr} which appears to
be instructive.
Introducing the notation
\[
{\rm B}_{\rm\sst in}(\kappa)\,\equiv\,\int\limits_{\BR_+}
d{\om}\;\langle \,T_{\rm\sst in}({\omega})\,
\,|\,B_{\kappa}\,\rangle\;T_{\rm\sst in}({\omega})
\]
allows us to write \rf{out->inappr} in the following form:
\begin{align}\label{out->inappr'}
& \langle\!\langle \,y\,|
\,u_-^\0\,\rangle\!\rangle_{}^{-1}
\langle\!\langle \,y\,|\,\rmb_{\rm\sst out}({\omega}_1)\cdots
\rmb_{\rm\sst out}({\omega}_n)\,|\,u_-^\0\,\rangle\!\rangle\,
\asymp_{g_s}\, \\
&\,\qquad\qquad\quad\asymp_{g_s}\,
\sum_{d=1}^{\infty}\,\frac{g_s^{d}}{d!} \; \big\langle
\,T_{\rm\sst out}({\omega}_1)\cdots T_{\rm\sst out}({\omega}_n)\,
({\rm B}_{\rm\sst in}(\kappa))^d\,\big\rangle^{}_{c=1}
\nonumber\\
&\,\qquad\qquad\quad\asymp_{g_s}\,
\big\langle \,T_{\rm\sst out}({\omega}_1)\cdots
T_{\rm\sst out}({\omega}_n)\,
\exp\!\big[{\rm B}_{\rm\sst in}(\kappa)\big]\,\big\rangle^{}_{c=1.}
\nonumber
\end{align}
It seems natural to call the representation \rf{out->inappr'} the
closed string representation. This representation suggests the
following interpretation in terms of 2d string theory. The initial
state of the D-brane is represented as a coherent state of
incoming closed strings\footnote{on top of an unobservable
D0-remnant, if you wish.}. The resulting out-state is obtained by
applying the closed string S-matrix to the closed string
oscillators which generate the initial state.

\subsection{Open string picture}\label{OSP}

There is an alternative way to calculate the amplitude
\rf{overlap1}. Let us first note the simple relation
\begin{equation}\label{ampl0}\begin{aligned}
\langle\!\langle \,y\,|\,
\rmb_{\rm\sst out}(\omega_1) & \dots
\rmb_{\rm\sst out}(\omega_n)\,
\,|\,F\,\rangle\!\rangle
\,= \,\\
=\,& \int_0^{\infty}du_+\;\chi_{\rm\sst out}(u_+^{})\prod_{r=1}^n
e^{i\de(\om_r)}\,e^{-i\om_r\log u_+}\,\lala\,y\,|\,u_+^{}\,\rr.
\end{aligned}\end{equation}
When calculating the matrix element $\lala\,y\,|\,u_+^{}\,\rr$
one should not forget that we had adopted the
convention to create the state $|\,y\,\rr$ with the help of the
fermionic in-field, cf. the footnote in subsection \ref{Fermidef}.
It follows that
\begin{equation}
\lala\,y\,|\,u_+^{}\,\rr\,\simeq_y\,(2y)^{-\frac{1}{2}}\,e^{i\mu\log
u_+} \,\rho^*(-\mu)\,u_+^{-\frac{1}{2}},
\end{equation}
where $\rho(\om)$ is the diagonal element of the single particle
reflection matrix defined in \rf{rhoom}.
We thereby arrive at the expression
 \begin{equation}\label{ampl1}\begin{aligned}
\langle\!\langle \,y\,|\,
\rmb_{\rm\sst out}(\omega_1) & \dots
\rmb_{\rm\sst out}(\omega_n)\,
\,|\,F\,\rangle\!\rangle
 \,\simeq_y \,(2y)^{-\frac{1}{2}}\rho^*(-\mu)\,
\tilde{\chi}_{\rm\sst out}^{\rm\sst R}(\omega-\mu)
\,\prod_{r=1}^{n}
e^{i\delta(\omega_r)}
\end{aligned}\end{equation}
where $\omega=\sum_{r=1}^{n}\omega_r$, and
$\tilde{\chi}_{\rm\sst out}^{\rm\sst R}(\omega)$ is the Fourier-transformation of $\chi_{\rm\sst out}^{\rm\sst R}
(u_+)$,
\[
\tilde{\chi}_{\rm\sst out}^{\rm\sst R}(\omega)=
\int_0^{\infty}du\;u^{-i\om-\frac{1}{2}}\,
\chi_{\rm\sst out}^{\rm\sst R}
(u).
\]
It remains to calculate
$\tilde{\chi}_{\rm\sst out}$
for the
choice of $\tilde{\chi}_{\rm\sst in}$
which corresponds to
the state $|\,u_-^\0\,\rangle\!\rangle$,
cf. \rf{chiin}, namely
\[
\bigg(\begin{matrix}
\chi_{\rm\sst in}^{\rm\sst L}(\om)\\ % [-1ex]
\chi_{\rm\sst in}^{\rm\sst R}(\om)
\end{matrix}
\bigg)\,=\,
e^{-i\om \log u_-^\0}\,\bigg(\begin{matrix}
0\\ % [-1ex]
1
\end{matrix}
\bigg).
\]
The relation \rf{oscrel} simplifies to $\chi_
{\rm\sst out}^{\rm\sst R}(\om)
=\rho(\om)\chi_{\rm\sst in}^{\rm\sst R}(\om)$.
We finally arrive at the simple result
\begin{equation}
\boxed{\quad\begin{aligned}
\langle\!\langle \,y\,|\, &
\rmb_{\rm\sst out}(\omega_1)  \dots
\rmb_{\rm\sst out}(\omega_n)\,
\,|\,F\,\rangle\!\rangle\,\simeq_y\\
& \simeq_y\,(2y)^{-\frac{1}{2}}\,
\prod_{r=1}^n
e^{i\de(\om_r)}\,
\rho^*(-\mu)\rho(\om-\mu)e^{-i(\om-\mu)\log u_-^\0}.\end{aligned}\quad }
\end{equation}

%The corresponding wave-function
%$\chi_{\rm\sst out}^{\rm\sst R/L}(\tau)$
%can then be calculated as follows:
%\begin{equation}\label{outRin}
%\tilde{\chi}_{\rm\sst out}^{}(\omega)\,=\,{\rm\mathbf R}(\omega)
%\cdot\tilde{\chi}_{\rm\sst in}^{}(\omega)\,,\quad
%{\rm\mathbf R}(\omega)=\Bigg(\begin{matrix}
%\rho(\omega) & \theta(\omega) \\
% \theta(\omega) & \rho(\omega)
%\end{matrix}
%\Bigg)\,
%\end{equation}
%where
%\begin{equation}
%\rho(\omega)=\frac{i}{\sqrt{2\pi}}e^{-\frac{\pi}{2}\omega}
%\Ga\big(\fr{1}{2}-i\omega\big),\qquad
%\theta(\omega)=ie^{\pi\omega}\rho(\omega),
%\end{equation}

\noindent{\bf Remarks}\\[1ex]
{\bf 1.} Viewing the fermionic field theory as a representation
for the open string theory on a gas of $D0$-branes \cite{McGV,KMS}
motivates us to call the resulting representation the ``open
string picture''. Quantum corrections to the D-brane dynamics are
calculated in the dual open string theory before we analyze the
final state in terms of closed string observables.

In the gedankenexperiment proposed in subsection \ref{interpret}
we are of course not tracking the evolution of the D-brane state
at finite times, we only observe outgoing radiation at late times.
It is therefore completely arbitrary\footnote{to the extend that
we can identify the different solitonic superselection sectors as
physically equivalent} if we prefer to interpret the state at
finite times as a D-brane or as a coherent state of closed
strings. We may in particular imagine that the D-brane is
"created" by an incoming coherent state of closed strings, and
that it subsequently decays back into an outgoing coherent state
of closed string radiation, as suggested by the "full-brane"
picture.
\\[1ex]
{\bf 2.} It is possible to calculate the expansion in powers of
$g_s$ by  noting that
\begin{equation}
\rho(\omega-\mu)\,=\,\frac{e^{-\frac{\pi}{2}(\om-\mu)}}{2\cosh\pi(\om-\mu)}
e^{i\xi(\omega-\mu)}
\,\underset{\mu\ra\infty}\asymp\,
e^{i\xi(\omega-\mu)}  ,  \quad
\xi(x)\equiv {\rm arg}\Ga\big(\fr{1}{2}-ix\big),\,
\end{equation}
for $\mu\ra\infty$ and $\frac{\om}{\mu}\ll 1$, and using
\begin{equation*}
e^{i\xi(\om-\mu)}
\asymp e^{i(\mu\ln\mu-\mu)}\,
\mu^{-i\om}\,
\exp\bigg(i\sum_{n=1}^{\infty}
\frac{(-1)^n B_{2n}}{2n(2n-1)}(1-2^{-(2n-1)})(\om-\mu)^{-(2n-1)}\bigg).
\end{equation*}
Note that we reproduce our
previous result \rf{onept-as}
for the leading asymptotics $\mu\ra\infty$, $\frac{\om}{\mu}\ll 1$ of the
one-point function.

The individual terms in the
resulting expansion are naturally interpreted as perturbative
contributions to the amplitude for emission of closed
strings in the time-dependent
background that is furnished by the decaying ZZ-brane,
corresponding to the following reorganization of the perturbative expansion
\rf{ampl-exp}:
\begin{equation}\label{discsum}% \begin{aligned}
\big\langle \,T_{\rm\sst out}({\omega}_1)\cdots  T_{\rm\sst
out}({\omega}_n)\, \big\rangle^{}_{\rm D0,\kappa}
=\sum_{r=0}^{\infty}\,g_s^{r-2}\, \sum_{\substack{h,d=0
\\2h+d=r}}^{\infty} \,\big\langle \,T_{\rm\sst
out}({\omega}_1)\cdots T_{\rm\sst
out}({\omega}_n)\,\big\rangle^{\sst(d,g)}_{c=1}\,.
%\end{aligned}
\end{equation}
One should bear in mind that the perturbative expansion in powers
of $g_s$ will give a useful approximation only if $\mu\ra\infty$,
and if only low energy tachyons with $\frac{\om}{\mu}\ll 1$ are
``measured''.\\[1ex]
{\bf 3.} It may seem puzzling that in \rf{u_in-kappa} we are
identifying the D-brane parameter with the initial localization
for time $t\ra-\infty$, whereas in \cite{KMS,DKKMMS} it is
associated with the turning point of the classical motion in the
inverted harmonic oscillator potential. By forming wave-packets it
is of course possible to get states which are in the classical
limit $\mu\ra\infty$ well-localized in the sense that the
uncertainties $\frac{1}{\mu}(\de u_\pm)^2$ and
$\frac{1}{\mu}(\de\la)^2$ are small. For those wave-packets one
recovers \rf{overlap} as the leading approximation to the tachyon
emission amplitude.

\subsection{More general $c=1$-backgrounds}

More generally we may consider string scattering amplitudes of the
form
\begin{equation}\label{backgrscatt}
\big\langle\!\big\langle \,f_+^{}\,|\,
\rmb_+({\omega}_1)\cdots
\rmb_+({\omega}_n)\,
\rmb_-(-{\omega}_1')\cdots
\rmb_-(-{\omega}_m')\,
|\,f_-^{}\,\big\rangle\!\big\rangle,
\end{equation}
where
\begin{equation}\label{cohst}\begin{aligned}
 |\,f_-\,\rangle\!\rangle\,=\,&
\exp\left(\int_{-\infty}^0 d{\om}\;f_-({\om})
\rmb_-({\om})\right)|\,\mu\,\rangle\!\rangle\\
\lala\,f_+\,|\,=\,&
\lala\,\mu\,|\,\exp\left(\int^{\infty}_0 d{\om}\;f_+({\om})
\rmb_-({\om})\right).
\end{aligned}\end{equation}
This may be interpreted as a string scattering amplitude in
a time-dependent background that is explicitly
represented in terms of  coherent states
$|\,f_+^{}\,\rangle\!\rangle$,
$|\,f_-^{}\,\rangle\!\rangle$ of closed strings. Of course it
is sufficient to study
\begin{equation}\label{genfct}
\langle\!\langle \,f_+^{}\, %{\mathsf S}\,
|\,f_-^{}\,\rangle\!\rangle,
\end{equation}
from which \rf{backgrscatt} can be recovered by taking functional
derivatives.

We now want to describe the scattering amplitudes in a background
that contains a decaying ZZ-brane on top of the closed string background
described by $|\,f_+^{}\,\rangle\!\rangle$ and
$|\,f_-^{}\,\rangle\!\rangle$. Our previous discussions suggest that
\begin{equation}
{}^{}_\1\langle\!\langle \,f_+^{}\,|\,
\Psi^{\dagger}_-(u_-^\0)\,
|\,f_-^{}\,\rangle\!\rangle
\end{equation}
represents the generating functional for the amplitudes in
question. The state ${}^{}_\1\langle\!\langle \,f_+^{}\,|$ is
defined by replacing the vacuum $\lala\,\mu\,|$ in \rf{cohst} by
an approximate vacuum in the one fermion sector. By using our
partial bosonization formula \rf{partbose} it becomes easy to show
that
\begin{equation}\label{backgrshift}\boxed{\begin{aligned}
{}^{}_\1\langle\!\langle \,f_+^{}\,|\,
 & \Psi^{\dagger}_-(u_-^\0)\,
|\,f_-^{}\,\rangle\!\rangle\,=\,\\
&=\, \exp\left(i\int_{\BR_+}d\om \,(u_-^\0)^{i\om}
f_{-}^{}(-\om)\right)\,
{\phantom{\big\rangle}}^{}_\1\big\langle\!\!\big\langle
\,f_+^{}\,|\, {\mathsf S}\,
|\,f_-^{}\!+e_\tau\,\big\rangle\!\!\big\rangle^{}_\1
\end{aligned}}\;\,,\end{equation}
where $e_{\tau}$
is explicitly given as
\[
e_{\tau}(\om)\,=\,i\,
\frac{(u_-^\0)^{i\omega}-e^{\omega y}}{\om}.
\]
Equation \rf{backgrshift} shows that the insertion of a decaying
ZZ-brane generates a shift in the closed string background that is
linear in the variables $f_+^{}$. The open-closed duality
expressed by equation \rf{backgrshift} is not perfect, though. The
infrared region near $\om=0$ is effectively removed by our cut-off
$y$. We therefore do not generate a shift of the cosmological
constant $\mu$, corresponding to the zero energy tachyon. The fact
that we can not remove the infrared cut-off $y$ limits the extend
to which strict open-closed duality is realized in our context. On
the other hand, our previous discussion of this issue shows that
what we are missing to strict open-closed duality is associated
with low-energy quanta that we are not able to observe anyway. In
this sense one may well regard the failure of strict open-closed
duality as unphysical.

\section{Comparison with the euclidean case}

It seems worth pointing out a close analogy between the results of the
previous subsection and discussions of the integrable
structure of two-dimensional string theory in
\cite{DMP},\cite{AKK} and \cite{ADKMV} respectively. In the following
we shall review those features of the formalisms developed
in  \cite{DMP,AKK,ADKMV} which we need to see the analogy with
the results in the previous section. Our discussion will not
be self-contained, the reader not sufficiently familiar with
the results of \cite{DMP,AKK,ADKMV} may need to consult these
references while reading the following section.

\subsection{Euclidean generating function}

We will now consider the case of euclidean target
space for the two-dimensional string theory
which is obtained by $X_0\ra -iX_0\equiv X$.
Compactification of euclidean time via
$X\equiv X+2\pi R$ will be introduced to describe
finite temperature. One is then in particular interested in
deformations of the background induced by changing the
world-sheet action as
\begin{equation}\label{WSdef}
S_{\rm WS} \rightarrow S_{\rm WS} + \sum_{k\neq 0}\, t_k \,
T_{\rm \sst E}(p_k),
\end{equation}
where $p_k=k/R$ and
$T_{\rm \sst E}(p)$ is the on-shell vertex operator
\[
T_{\rm \sst E}(p)\sim \int d^2z \;e^{ipX}e^{2(1-|p|)\phi}\,.
\]
The central object to study is the deformed partition
function
\begin{equation}\label{euclpart}
Z(\{t_k\};\mu,R)\,=\,\bigg\langle\exp\Big(
-\sum_{k\neq 0}\, t_k \,
T_{\rm \sst E}(p_k)\Big)\bigg\rangle_{\rm c=1.}^{\rm Eucl}
\end{equation}

Turning to the free fermionic field theory,
it is in fact straightforward to introduce a natural
euclidean counterpart to \rf{genfct} as follows \cite{MPR}:
Let $u_{\pm}=e^{t\mp x}$ and continue $t=i\theta$.
Periodicity w.r.t. $\theta\ra\theta+2\pi R$ then leads to
quantization of the euclidean energies as $\om=ik/R$.
The euclidean
counterparts of in- and out bosonic fields,
\begin{equation}
\pa_{u_\pm}S_\pm(u_\pm)=\sum_{k\in\BZ}a_k^{(\pm)}
u_{\pm}^{-\frac{1}{R}(k+1)},\quad \pm=\pm
\end{equation}
will be single-valued. It is then natural
to consider
\begin{equation}\label{euclZ}
\CZ(\{ t_k\};\mu,R)\,=\,\lala \,T_+\,|\, T_-\,\rr,
\end{equation}
where $|T_-\rr$, $\lala T_+|$ are coherent states of
bosonic excitations defined as
\begin{equation}
|\, T_+\,\rr\,\equiv\,\exp\bigg(
\sum_{k<0}t_k a_k^{\rm\sst in}\bigg)\,|\,\mu\,\rr, \qquad
\lala \,T_-\,|\,\equiv
\lala\,\mu\,|\exp\bigg(\sum_{k>0}
t_k a_k^{\rm\sst out}\bigg).
\end{equation}
The generating function \rf{euclZ} can be evaluated by first
expressing the bosonic oscillators $a_k^{(\pm)}$ in terms
of fermions as
\begin{equation}
a_k^{(\pm)}\,=\,\sum_{l\in\BZ+\frac{1}{2}}\rmd_k^{(\pm)}
\rmd_{k-l,}^{\dagger(\pm)}\qquad
\begin{aligned}
&\psi_{\pm}(u_\pm)=\sum_{k\in\BZ+\frac{1}{2}}\rmd_k^{(\pm)}
u_{\pm}^{-\frac{k}{R}-\frac{1}{2R}},\\
&\psi_{\pm}^{\dagger}(u_\pm)=\sum_{k\in\BZ+\frac{1}{2}}
\rmd_k^{\dagger(\pm)}
u_{\pm}^{-\frac{k}{R}-\frac{1}{2R}},
\end{aligned}\end{equation}
and then using the algebra
$\big\{\rmd_k^{(\pm)},\rmd_l^{\dagger(\pm)}\big\}=\de_{k+l}$
as well as the following relation between in- and
out oscillators:
\begin{equation}
\rmd_k^{\rm\sst out}\,=\,\rho(p_k)\,\rmd_k^{\rm\sst in}.
%\qquad \rmd_k^{\dagger\rm\sst out}\,=\,(\rho(p_k))^*\,
% \rmd_k^{\dagger\rm\sst in}
\end{equation}
Equation \rf{euclZ} is therefore good enough to define
$\CZ$ as a formal series in the variables $t_k$.
The conjectured duality between the euclidean versions
of $c=1$ string theory and free fermionic field theory
is coincidence of the objects defined in
\rf{euclZ} and \rf{euclpart}, respectively.

\subsection{Deformations of the Fermi level curve}

The starting point of the formalism developed in \cite{AKK} is the
reformulation of the free fermionic field theory in terms of
light-cone variables $u_\pm$ for the phase space of the single
particle problem,
\begin{equation}
u_\pm\equiv\frac{1}{{2}} \la\pm p.
\end{equation}
The classical single particle
Hamiltonian is then simply $h=-u_+u_-$, so that the vacuum of the
classical limit of free fermionic field theory, the filled
Fermi sea, gets represented by the equation
\begin{equation}\label{glue}
u_+u_-=\mu.
\end{equation}
Complexifying $u_\pm$ one may regard \rf{glue} as the definition
for a noncompact Riemann surface which may be covered by two
patches $\CU_\pm$ with coordinates $u_\pm$ respectively. Eqn.
\rf{glue} defines the transition between the patches $\CU_+$ and
$\CU_-$.

In the quantized theory $u_\pm$ get represented by the
operators
\[ \su_\pm\,=\,\frac{1}{2}\la\mp i\pa_\la.
\]
One may introduce representations for the single particle
Hilbert space in which either $\su_+$ or $\su_-$ are
diagonal. The representation of eigenfunctions of the
hamiltonian $\sh=-\su_+\su_--\su_-\su_+$ becomes very simple,
\begin{equation}\label{lighteigen}
\zeta^{s}_{\pm}
(\om|u_\pm)\,=\,
\frac{1}{\sqrt{2\pi}}
\Theta(s u_\pm)
|u_{\pm}|^{\pm i\om-\frac{1}{2}}\,.
\end{equation}
Thanks to the fact that $[\su_+,\su_-]=-i$ one may realize the
unitary operator between these two representations simply as
Fourier transformation,
\begin{equation}\label{FT}
\phi(u_+)=\frac{1}{\sqrt{2\pi}}\int du_-\;e^{-iu_+u_-}\phi(u_-).
\end{equation}

The relation between the formalism introduced in this paper and
the light-cone formalism of \cite{AKK} follows from the
fact that the unitary transformation from wave-functions
$\psi(\la)$ to their time asymptotics $\phi_{\pm}(u_\pm)$
diagonalizes the operators $\su_\pm$. This is proven
in appendix \ref{aswave}.

It is not hard to set up a formalism for the
second quantized theory in terms of the variables $u_\pm$.
Key ingredients of this formalism will be the
fermionic field operators $\Psi_{\pm}(u_\pm)$.
It is useful to associate the two field operators
with the corresponding patches $\CU_\pm$. It follows from
\rf{FT} that the operators $\Psi_{+}(u_+)$
and $\Psi_{-}(u_-)$
are also related by Fourier transformation,
\begin{equation}\label{FermiFou}
\Psi_{+}(u_+)=
\frac{1}{\sqrt{2\pi}}\int du_-\;e^{iu_+u_-}\Psi_{-}^s(u_-).
\end{equation}

In order to treat
deformed backgrounds of the
euclidean along the lines of \cite{AKK} one may start from
the following key idea:
{\it The deformations can be described
by a change of the vacuum in which to calculate
expectation values
{\it only}}.
The deformation should therefore not change any of the
relations which characterize the operator algebra of the
theory including the relation
between in- and out fields \rf{FermiFou}.

The deformation will induce, however,
a  deformation of the energy eigenfunctions \rf{lighteigen}
which appear in the expansion of the Fermi-fields
into creation- and annihilation operator with a
specific energy. The authors of \cite{AKK} propose that the
deformation of the energy eigenfunctions will take the
form\footnote{For the ease of notation we restrict ourselves
to $u_\pm>0$ in the following.}
\begin{equation}\label{lighteigendef}
\zeta_{\pm}^{\theta}
(\om|u_\pm^{})\,=\,
\frac{1}{\sqrt{4\pi}}\,e^{i\theta_\pm(\om|u_\pm)}
u_{\pm}^{\pm i\om-\frac{1}{2}}\,,
\end{equation}
with phases $\theta_\pm(\om|u_\pm)$ that are of the form
\begin{equation}
\theta_\pm(\om|u_\pm)\,=\,\frac{1}{2}\theta_0(\om) +
\sum_{k\geq 1} \,t_{\pm k}\,u_{\pm}^{\frac{k}{R}}
-\sum_{k\geq 1} \frac{1}{k}v_{\pm k}(\om)
u_{\pm}^{-\frac{k}{R}}.
\end{equation}
A basic idea behind this proposal is that the ``field''
$\theta_\pm(\om|u_\pm)$ should essentially coincide with
the expectation value of the bosonic fields $S_\pm(u_\pm)$
obtained by bosonizing the fermionic fields $\Psi_{\pm}(u_\pm)$.
More precisely, the relation with the deformed
partition  function $\CZ=\CZ(\{t_k\};\mu,R)$
is expected to be
\begin{equation}\label{thetaexp}
\theta_\pm(\om|u_\pm)\,=\,\frac{1}{\CZ}\biggl(
\frac{1}{2}\frac{\pa}{\pa \mu} +
\sum_{k\geq 1} \,t_{\pm k}\,u_{\pm}^{\frac{k}{R}}
-\sum_{k\geq 1} u_{\pm}^{-\frac{k}{R}}
\frac{1}{k}\frac{\pa}{\pa t_{\pm k}}\bigg)\CZ.
\end{equation}
In order for \rf{FermiFou} to remain valid in the deformed
theory one then needs that the deformed energy eigenfunctions
are related by
\begin{equation}\label{defFou}
\zeta_{+}^{\theta}
(\om|u_+^{})\,=\, \frac{1}{\sqrt{2\pi}}\int du_-\;e^{-iu_+u_-}
\zeta_{-}^{\theta}
(\om|u_-^{})
\end{equation}
This defines an intricate problem. Regarding the coefficients
$t_k$ as given input data one finds from \rf{defFou} a non-trivial
set of relations between the parameters $t_k$ and the coefficients
$v_{\pm k}(\om)$.  One may expect that these relations can
generically be solved to uniquely to define $v_{\pm k}(\om)$ as a
function of the $t_k$. The deformed partition function $\CZ$ is
then defined via \rf{thetaexp}, where the integrability of these
equations follows from the observation that a solution to this
problem defines a particular solution of the Toda integrable
hierarchy \cite{AKK}.

All this can be understood much more concretely in the classical
limit $\mu\ra\infty$. In this case one may evaluate \rf{defFou}
via the saddle point method \cite{AKK}, leading to the conditions
\begin{equation}\label{gluedef}
u_+u_-=\left\{
\begin{aligned}u_+\pa_{+}S_+(u_+)\equiv
&\,\mu+\sum_{k\geq 1}\,k\,t_{+k}\,u_+^{\frac{k}{R}}\,+\,
\sum_{k\geq 1}\,v_{+k}\,u_+^{-\frac{k}{R}}\\
u_-\pa_{-}S_-(u_-)\equiv&\,\mu+ \sum_{k\geq
1}\,k\,t_{-k}\,u_-^{\frac{k}{R}}\,+\, \sum_{k\geq
1}\,v_{-k}\,u_-^{-\frac{k}{R}}\,.
\end{aligned}
\right.
\end{equation}
The coefficients $v_k$ are now
defined as functions of the $t_k$ by the mutual
consistency of the
two equations in \rf{gluedef}, see \cite{AKK} for details.
Having chosen the $v_k$ in such a way that
the equations \rf{gluedef} are consistent one may
view either of these equations as the defining equation
for a Riemann surface that is obtained as a deformation
of the surface \rf{glue}.

It seems worth remarking that the
corresponding classical free energy $\CF_{\rm cl}$,
defined by
\begin{equation}
v_k=-\frac{\pa}{\pa t_k} \CF_{\rm cl}\,,
\end{equation}
defines a natural K\"ahler potential,
whose associated symplectic form identifies
the coefficients $v_k$ as the dual momenta to the
coordinates $t_k$ for the space of deformations of the
surface \rf{glue}.
This line of thought naturally leads
to the proposal that the partition
function $\CZ$ of the quantized theory can be interpreted
as the wave-function of a particular state in the
quantization of the symplectic space with
K\"ahler potential $\CF_{\rm cl}$.
The Fourier transformation
\rf{FermiFou} may then be regarded
as the natural quantum counterpart
of the transition between the patches $\CU_+$ and
$\CU_-$. This point of view is strongly supported by the
observation from \cite{AKK}
that the Fourier transformation
\rf{FermiFou} reduces to \rf{gluedef} in the
classical limit.

A very similar framework was shown in \cite{ADKMV} to follow
from a general formalism for solving the topological
B-model on certain classes of noncompact Calabi-Yau
manifolds. The case of the $c=1$ string corresponds
to the hypersurface
\[
zw-H(p,\la)=0, \quad H(p,\la)=p^2-\frac{1}{4}\la^2-\mu.
\]
In this context one interprets the fermionic fields
$\Psi_\pm(u_\pm)$ as representatives for topological D-branes that
may be present in the relevant Calabi-Yau geometry. These branes
are parameterized by points of the surface $H(p,\la)=0$, or
alternatively by the corresponding values of the coordinates
$u_\pm$.

\subsection{Comparison}

Although this has not been shown non-perturbatively yet, it seems
very likely that the formalisms outlined in the previous two
subsections are ultimately all equivalent. One way to establish
this is to observe that all these formalisms produce solutions of
the equations of the Toda hierarchy with initial conditions given
by the partition function of the undeformed two-dimensional string
theory background.

In any of these formalisms an important role is played by
the one-point functions of the fermionic fields,
which will be denoted as
\begin{equation}
\big\langle\,\Psi_\pm(u_\pm)\,\big\rangle_{\{t_k\};\mu,R.}^{}
\end{equation}
By using standard bosonization formulae it is then
straightforward to show that e.g.
\begin{equation}\label{euclbackgrshift}
\big\langle\,\Psi_-(u_-)\,\big\rangle_{\{t_k\};\mu,R}^{}
=\exp\Bigg(\sum_{k\geq 1} \frac{1}{k}\,t_{k}\,
u^{-\frac{k}{R}}_{-}
\Bigg)\CZ\Big(\,\Big\{\,t_k+\frac{i}{k}u_-^{\frac{k}{R}}\Theta(-k)
\,\Big\}\,;\,\mu,R\,\Big).
\end{equation}
Following \cite{ADKMV} one may read this as follows:
{\it Insertion of a topological B-brane at position $u_-$
generates the shift \[
t_k\ra +\frac{i}{k}\,u_-^{\frac{k}{R}}\,\Theta(-k)\]
of the closed string background.}

The analogy between \rf{euclbackgrshift} and
\rf{backgrshift} should be clear. But our discussion
also shows that the relation between \rf{euclbackgrshift} and
\rf{backgrshift} is more than just an analogy: Bear
in mind that the coordinates $u_\pm$ with which
we describe the in- and out states are {\it identical} with the
light cone coordinates which play a central role in the
euclidean formalisms. It follows that the
fermionic fields  $\Psi_\pm(u_\pm)$ of these
formalisms are nothing but the euclidean counterparts of
the fermionic in- and out fields in the
Minkowskian formalism used in this paper. One of our main
results is to show that the insertion of
fermionic in-fields describes decaying ZZ-branes.
This finally leads us to propose
that {\it the topological B-branes of \cite{ADKMV} are the
euclidean counterparts of the rolling ZZ-branes.}

\section*{Acknowledgements}
This work has its roots in discussions with S.-J. Rey. I gratefully
acknowledge these discussions, as well as collaboration in the
initial stages.

I am grateful to I. Klebanov, D. Kutasov, J. Maldacena, G. Moore,
R. Roiban, A. Schwimmer, H. Verlinde and J. Walcher for
interesting discussions related to this work. It is furthermore a
pleasure to thank the following institutions, where part of the
work was carried out, for hospitality: IAS Princeton, ESI Vienna,
Caltech, UC Berkeley, Univ. Rutgers and the Weizmann Institute,
Rehovot.

Last but not least I acknowledge financial support by
a Heisenberg fellowship of the DFG, as well as partial support
by the EC via the EUCLID research training network contract
HPRN-CT-2002-00325.

\newpage

\appendix
\section{Free fermionic field theory revisited}

\subsection{Single particle quantum mechanics}

As emphasized in \cite{AKK}, it is convenient to start by representing the
single particle Hamiltonian $\sh$ in terms of the
``light-cone variables'' $\su_\pm\,=\,\frac{1}{2}\la\pm i\pa_\la$,
\[
\sh=-\su_+\su_--\su_-\su_+.
\]
There exist representations for
the Hilbert space $\CK$ of the single particle problem in which
either $\su_+$ or $\su_-$ are represented as multiplication
operators. Observing that $\sh$ is the generator of dilatations
of the coordinates $u_\pm$
it becomes easy to find a complete set of eigenfunctions for
$\sh$,
\begin{equation}\label{lighteigenA}
\zeta^{s}_{\pm}
(\om|u_\pm)\,=\,
\frac{1}{\sqrt{2\pi}}
\Theta(s u_\pm)
|u_{\pm}|^{\pm i\om-\frac{1}{2}}\,,\quad s\in\{+,-\},~~\om\in\BR.
\end{equation}
These representations are related to the usual
Schr\"odinger representation by means of integral transformations
of the form
\begin{equation}\label{inttrsfA}
\phi_{\pm}^{}(u_\pm^{})\,=\,
\int_{\BR}d\la\;{\rm M}_{\pm}^{}(u_\pm^{}|\la)\,
\psi(\la),
\end{equation}
with kernels ${\rm M}_{\pm}(u_\pm|\la)\equiv\langle u_\pm|\la\rangle$
given by the explicit
formulae
\begin{equation}\label{intkernA}
{\rm M}_{+}(u_+|\la)\,=\,
e^{-i\frac{\pi}{4}}e^{\frac{i}{2}u^2_+-i\la u_++\frac{i}{4}\la^2},
\quad
{\rm M}_{-}(u_-|\la)\,=\,\big({\rm M}_{+}(u_-|\la)\big)^*_.
\end{equation}
This claim is easily verified using
the fact that the kernels ${\rm M}_{\pm}(u_\pm|\la)$
satisfy the differential equations
\[
\big(\pm i\pa_\la+\fr{1}{2}\la\big){\rm M}_{\pm}(u_\pm|\la)\,=\,u_\pm
{\rm M}_{\pm}(u_\pm|\la).
\]

When working in the Schr\"odinger representation one may construct
a convenient set of eigenfunctions for the single particle
hamiltonian $\sh$ by applying the inverse of the transformation \rf{inttrsfA}
to the eigenfunctions \rf{lighteigenA}. In this way one may construct
in particular
\begin{equation}
G(\om|\la)\,=\, \frac{e^{-\frac{\pi}{2}\om-i\frac{\pi}{4}}}
{\Ga\big(\frac{1}{2}-i\om)}\,
e^{i\frac{\la^2}{4}}
\int\limits_0^{\infty+i\ep}d\si \;\si^{-i\om-\frac{1}{2}}
\,e^{i\la\si+\frac{i}{2}\si^2}.
\end{equation}
$G(\om|\la)$ is an eigenfunction of $\sh$ with eigenvalue
$\om$ which has particularly simple asymptotics for $\la\ra+\infty$,
namely
\begin{equation}\label{Gasym}
G(\om|\la)\,\underset{\la\ra\infty}{\sim}\,
e^{i\frac{\la^2}{4}}\la^{i\om+\frac{1}{2}}.
\end{equation}
The functions $G(\om|\la)$ are related to the
standard parabolic cylinder functions $U(a,x)$ \cite{AS} via
\begin{equation}
G(\om|\la)=\,  e^{-\frac{\pi}{4}\om-i\frac{\pi}{8}}\,
U\big(-i\om,\la e^{-i\frac{\pi}{4}}\big).
\end{equation}
Three further solutions with simple asymptotic behavior
can be obtained as
$G(\om|-\la)$, $G^*(\om|\la)$, $G^*(\om|-\la)$, where the asterisk
denotes complex conjugation. A normalized set of
real parity eigenfunctions is finally constructed as
\begin{equation}\label{Fdef}
{\rm F}_p(\om|\la)\equiv\frac{1}{\sqrt{2\pi}}
\Big(m_p(\om)G(\om|\la)+m_p^*(\om)G^*(\om|\la)\Big)\,,
\end{equation}
where the coefficients $m_p(\om)$ are defined as
\begin{equation}
{\rm m}_p(\om)\,\equiv\,
\frac{e^{i\frac{\pi}{4}}}{\sqrt{2}}
\frac{k(\om)-ip}{\sqrt{k^2(\om)+1}}
\left(\frac{\Gamma(\fr{1}{2} - i \omega )}
{\Gamma(\fr{1}{2} + i \omega )}\right)^{\frac{1}{4}},
\end{equation}
with
$k(\omega) = \sqrt{1 + e^{ - 2 \pi \omega} } - e^{ - \pi \omega}$.
The label $p=\pm$ is identified with
the parity eigenvalue of ${\rm F_p}(\om|\la)$.
The functions ${\rm F_p}(\om|\la)$ have asymptotics
\begin{equation}\label{Fasym}
{\rm F}_p(\om|\lambda)
\,\underset{|\la|\ra\infty}{\sim}\,
\frac{1}{\sqrt{2\pi |\lambda|)}}
\Big(e^{\frac{i}{4}\la^2}\,e^{i\om\ln|\la|}\,{\rm M}_p^s(\om)
+e^{-\frac{i}{4}\la^2}\,e^{i\om\ln|\la|}\,
\overline{\rm M}{}_{p}^s(\om)\Big),
\end{equation}
where ${\rm M}_p^s(\om)= s^{\Theta(-p)}{\rm m}_p(\om)$ with
$\Theta(-p)$ being the usual step function. It is known \cite{M}
that the functions ${\rm F_p}(\om|\la)$ fulfil the following
orthogonality and completeness relations :
\begin{equation}\begin{aligned}
&{}\int_\BR d\la\;{\rm F}_{p_\1}(\om_\1|\la)
{\rm F}_{p_\2}(\om_\2|\la)\;=\;\de_{p_\2p_\1}\,\de(\om_\1-\om_\2),\\
&  \int_\BR d\om \;\big(\, {\rm F}_{}^+(\om|\la_\1)
{\rm F}_{}^+(\om|\la_\2)+{\rm F}_{}^-(\om|\la_\1)
{\rm F}_{}^-(\om|\la_\2)\,\big)\;=\;\de(\la_\1-\la_\2).
\end{aligned}\end{equation}

\subsection{Asymptotics of wave-packets} \label{aswave}

{\bf Claim:} {\it The asymptotics of a wave-packet $\psi(\la,t)$
for $t\ra\pm\infty$ is of the form
\begin{equation} \psi(\lambda, t) \quad
{}_{\widetilde{t \rightarrow \pm\infty}}
\quad (2 \pi)^{-\frac{1}{2}} e^{\frac{i}{4} \lambda^2} \,
e^{\mp\frac{t}{2}}\,\phi_{\pm}(u_\pm),
%\qquad\begin{aligned}& s\equiv \sgn(\la),\\
%& x\equiv\ln|\la|,\end{aligned}
\label{asymptA} \end{equation}
where $u_{\pm}\equiv \la e^{\mp t}$.
The  asymptotic wave-functions $\phi_{\pm}(u_\pm)$ can be
calculated from the wave function $\psi(\la)\equiv \psi(\la,0)$
by means of the integral transformations \rf{inttrsfA}.
}\\[1ex]
{\bf Proof:}
We may represent $\psi(\lambda, t)$ as
\begin{equation}
\psi(\lambda, t)\,=\,\int d\om \;e^{-i\om t}\,
{\mathbf F}(\omega|\lambda)\cdot\tilde{\psi}(\om),\quad
\tilde{\psi}(\omega) \equiv
\int_{\mathbb{R}} \,\rmd \lambda \, {\mathbf F}(\omega|\lambda)
\psi(\lambda),
\end{equation}
Standard stationary phase arguments show that $\psi(\lambda, t)$ will
vanish rapidly at any fixed $\la$ when $|t|\ra\infty$.
We should therefore
regard the asymptotics where $|\la|$ tends to $\infty$ as well.
In this case we may replace the wave-functions
${\mathbf F}(\omega|\lambda)$
by their leading asymptotics for $|\la|\ra\infty$ as given
in equation \rf{Fasym}.

Only the term containing the factor $e^{-i\om(t\mp \ln|\la|)}$
will contribute in the limit $t\ra
\pm\infty$. This is enough to establish the first half
of our claim, equation \rf{asymptA}, with the two-component vector
$\phi_{\pm}(u_\pm)$ formed out of the functions
$\phi_{\pm}^s(u_\pm)$, $s=\pm$ given by
\begin{equation}\label{phiasFourier}
\phi_{\pm}(u_\pm)
\,=\,\int_\BR d\om \;u_{\pm}^{\pm i\om}
% e^{-i\om t}
\,{\mathbf M}_{\pm}(\om)\cdot
\tilde{\psi}(\om)\,,
\quad\begin{aligned}&
{\mathbf M}_{+}(\omega)\equiv {\mathbf M}(\omega),\\
& {\mathbf M}_{-}(\omega)\equiv \overline{\mathbf M}(\omega),
\end{aligned}
\end{equation}
where ${\mathbf M}(\omega)$ is the matrix with matrix elements
${\rm M}_p^s(\om)$, or explicitly
 \begin{equation}\label{Mmatrrepr}
{\rm\mathbf M}(\omega)\,\equiv\,\frac{1}{2}\bigg(\,\begin{matrix}
1& \phantom{-}1\\ % [-1ex]
1 & -1
\end{matrix}\,\bigg)\bigg(\begin{matrix}
{\rm m}_+(\omega) & 0\\ % [-1ex]
0 & {\rm m}_-(\omega)
\end{matrix}\bigg) .
\end{equation}
It remains to calculate the asymptotic wave-functions
$\phi_{\pm}^s(t)$ more explicitly.
To this aim let us consider ${\rm\mathbf M}(\omega)\cdot
{\mathbf F}(\om|\la)$. By using \rf{Fdef}, the
expression that follows from \rf{Fdef} by
$F_p(\om|-\la)=(-)^{\Theta(-p)}F_p(\om|\la)$ as well
as $|m_p(\om)|^2=\frac{1}{2}$ one arrives
at
\[ {\rm\mathbf M}(\omega)\cdot
{\mathbf F}(\om|\la)\,=\,\frac{1}{2\sqrt{2\pi}}
\big(m_+^2(\om)-m_-^2(\om)\big)\biggl(\begin{matrix}
G(\om|-\la) \\
G(\om|+\la)
\end{matrix}
\biggr).
\]
The factor $m_+^2(\om)-m_-^2(\om)$ equals
$2(2\pi)^{-\frac{1}{2}}e^{\frac{\pi\om}{2}}\Gamma\big(\frac{1}{2}-
i\om\big)$, leading us to
\[
{\rm\mathbf M}(\omega)\cdot
{\mathbf F}(\om|\la)\,=\,\frac{1}{2\pi}e^{-i\frac{\pi}{4}
+\frac{i}{4}
\la^2}\int\limits_0^{\infty+i\ep}d\si\; \si^{-i\om-\frac{1}{2}}
e^{\frac{i}{2}\si^2}\biggl(\begin{matrix}e^{-i\la\si}\\
e^{+i\la\si}\end{matrix}\biggr).
\]
This should then be inserted into \rf{phiasFourier}.
After exchanging
the integrations one can easily do the integration over
$\om$, thereby producing a delta-function. This straightforwardly
yields our claim that $\phi_\pm$ are given by the
integral transformation \rf{inttrsfA}.
\hfill$\square$

\subsection{In- and Out-fields}\label{AppIO}

Our next aim is to study
the asymptotics for $t\ra\pm\infty$ of the fermionic operators
\begin{equation}
\Psi^{\dagger}[\,\psi\,|\,t\,)\,\equiv\,
\int d\lambda\; \psi(\la)\,\Psi^{\dagger}(\la,t)\,.
\end{equation}
Introducing the operators ${\mathbf d}^{\dagger}_{\pm}(\om)$
by
${\mathbf d}^{\dagger}_{\pm}(\om)=
{\mathbf M}_{\pm}^{\dagger}
(\omega)\cdot
{\mathbf c}^{\dagger}(\om)$
allows us to write
\begin{equation}
\begin{aligned}
\Psi^{\dagger}[\,\psi\,|\,t\,)\,=\,& \int_\BR d\om\;e^{-i\om t}\,
{\mathbf c}^{\dagger}(\om)\cdot \tilde{\psi}(\om)\,\\
\,=\,& \int_\BR d\om\;e^{-i\om t} \,
{\mathbf d}^{\dagger}_{\pm}(\om)\cdot \tilde{\phi}_{\pm}(\om),
\end{aligned}\end{equation}
where the definition of $\tilde{\phi}_{\pm}(\om)$ can be read off from \rf{phiasFourier}.
Using \rf{umodes}
it is then straightforward to rewrite the result in the following form:
\begin{equation} \Psi^{\dagger}[\,\psi\,|\,t\,)
\,=\, \int \frac{du}{2\pi}\; \phi_{\pm}(u e^{\mp t})\,
\Psi^{\dagger}_{\pm}(u) \,.
\end{equation}
This clearly identifies the fermionic fields
$\Psi^{\dagger}_{\pm}(u)$ as the in- and out-fields.

\subsection{Relation between bosonic in- and out-oscillators}\label{inoutbose}

We want to demonstrate the validity of the expansion
\begin{align}\label{oscexpA}
[\,\rma_{\rm out}^{s}(\omega)\,]_{\0,\rm in}^{}\,
=\sum_{s'=\rm\sst L,R}\, \sum_{n=1}^{\infty}
\int\limits_{-\infty}^{\infty} &
\frac{d\om_1}{\om_1}\int\limits_{\om_1}^{\infty}\frac{d\om_2}{\om_2}
\cdots
\int\limits_{\om_{n-1}}^{\infty}\frac{d\om_n}{\om_n}\\
& \ti \;R_{(n)}^{ss'}(\,\om\,|\,\om_1,\dots,\om_n\,)\,
\rma_{\rm in}^{s'}(\omega_1)\cdots
\rma_{\rm in}^{s'}(\omega_n)\,.\nonumber
\end{align}
This formula is a nonperturbative generalization of Polchinski's
result \cite{Po2} for the classical limit $\mu^{-1}\ra 0$. The
coefficients $R_{(n)}^{ss'}$ are of the form $R_{(n)}^{ss'}
(\om|\om_1,\dots,\om_n)= 2\pi \de(\om-\sum_{r=1}^s\om_s)
Q_{(n)}^{ss'}  (\om_1,\dots,\om_n),$ where
\begin{equation}\label{Rdefs}\begin{aligned}
& Q_{(n)}^{ss'}(\,\om_1,\dots,\om_n\,)\,= \,
\int\limits_{-\infty}^{\infty}
\frac{d\tau}{i\tau}\;e^{-i\mu\tau}K^{ss'}(\om|\tau) \prod_{r=1}^n
2i\sin\fr{\om_r \tau}{2}.
\end{aligned}\end{equation}
The
kernels $K^{ss'}(\om|t)$ are given by the following integrals:
\begin{equation}
K^{ss'}(\,\om\,|\,t\,)=\!\!\int\limits_{-\infty}^{\infty}
d\om'\;e^{i\om't}\, \bar{\rm R}^{ss'}(\fr{\om'-\om}{2})
     {\rm R}^{ss'}(\fr{\om'+\om}{2})=\left\{
\begin{aligned}
&
J_{-i\om}(+2ie^{t})~~{\rm if}~~s=s',\\
& J_{-i\om}(-2ie^{t})~~{\rm if}~~s\neq s'\,.
\end{aligned}\right.
\end{equation}
{\bf Proof of \rf{oscexpA}}: We start from the expression
\begin{equation}%\begin{aligned}
{[}\,\rma_{\rm out}^{s}  (\omega)\,{]}_{\0,\rm in}^{}\,=\,
2\int\limits_{-\infty}^{\infty} d\om'\sum_{s'=\rm\sst L,R}
\bar{R}^{ss'}(\fr{\om'-\om}{2})
      R^{ss'}(\fr{\om'+\om}{2})\,
{\rm d}^{\dagger\,s'}_{\rm in}(\fr{\om'-\om}{2})
{\rm d}^{s'}_{\rm in}(\fr{\om'+\om}{2}).
%\end{aligned}
\end{equation}
By inserting
\[
{\rm d}^{\dagger\, s'}_{\rm
in}(\om)=\int\frac{dx_\1}{2\pi}\,e^{i\om x_\1}
\bar{\Psi}^{\dagger\, s'}_{\rm in}(x_\1),\quad {\rm d}^{s'}_{\rm
in}(\om)= \int\frac{dx}{2\pi}\,e^{i\om x_\2} \bar{\Psi}^{s'}_{\rm
in}(x_\2)\,,
\]
where $\bar{\Psi}^s_{\rm in}(x)\equiv e^{\frac{x}{2}}\Psi^s_{\rm
in}(e^x)$, exchanging the order of integrations and changing
variables to $x_\1=x+\frac{\tau}{2}$, $x_\2=x+\frac{\tau}{2}$, we
arrive at the expression
\begin{equation}\label{inter1}\begin{aligned}
{[}\,\rma_{\rm out}^{s} & (\omega)\,{]}_{\0,\rm in}^{}\,=\,\\
& =\,  \int\frac{dx}{2\pi}\;e^{-i\om x}\int \frac{d\tau}{2\pi}
\sum_{s'=\rm\sst L,R}
K^{ss'}(\,\om\,|\,\tau\,)\,
\Psi^{\dagger\, s'}_{\rm in}(x+\fr{\tau}{2})
\Psi^{s'}_{\rm in}(x-\fr{\tau}{2}),
\end{aligned}\end{equation}
where the kernel
$K^{ss'}(\,\om\,|\,\tau\,)$ is the one defined in \rf{Rdefs}.
For the product of fermionic field operators which appears in
\rf{inter1} we may use the bosonization formula
\begin{equation}\label{inter2}
\begin{aligned}
\Psi^{\dagger\, s}_{\rm in}  (x+\fr{\tau}{2}) &
\Psi^{s}_{\rm in}(x-\fr{\tau}{2})\,=\,\\
& =\,\frac{1}{i\tau}e^{-i\mu\tau}
e^{i(T_{<}^{s}(x+\frac{\tau}{2})-T_{<}^{s}(x-\frac{\tau}{2}))}
e^{i(T_{>}^{s}(x+\frac{\tau}{2})-T_{>}^{s}(x-\frac{\tau}{2}))}
\end{aligned}
\end{equation}
Note that this formula is free from the infrared problems of the
corresponding bosonization formula for a single fermionic field
operator. It may be proven by using the partial bosonization
formulae from \S\ref{Spartbose}, noting that the cut-off $y$ may
be removed in expressions that are bilinear in fermionic fields.

By using series
expansions for
the exponentials in \rf{inter2} we then get
\begin{equation}\label{inter3}
\begin{aligned}
& \Psi^{\dagger\, s}_{\rm in}  (x+\fr{\tau}{2})
\Psi^{s}_{\rm in}(x-\fr{\tau}{2})\,=\,\\
& =\,\frac{e^{-i\mu\tau}}{i\tau}\,\sum_{k=0}^{\infty}\,
\sum_{l=0}^k\frac{(2i)^k}{l!(k-l)!} \int\limits_{-\infty}^{\infty}
\frac{d\om_1}{\om_1}\cdots\frac{d\om_k}{\om_k}\,e^{i\bar{\om}x}\,
\prod_{r=1}^k \sin\fr{\om_r\tau}{2}\; \\
&\hspace{4.5cm} \times
{\rm a}_{\sst\rm in}^{s-}(\om_1)\cdots {\rm a}_{\sst\rm in}^{s-}(\om_l)
{\rm a}_{\sst\rm in}^{s+}(\om_{l+1})\cdots{\rm a}_{\sst\rm in}^{s+}(\om_{k}),
\end{aligned}
\end{equation}
where we have used the notation $\bar{\om}=\sum_{r=1}^k\om_r$
as well as ${\rm a}_{\sst\rm in}^{s\pm}(\om)\equiv\Theta(\pm\om)
{\rm a}_{\sst\rm in}^{s}(\om)$.
Passing to integration over ordered sets of integration
variables finally yields the
formula
\begin{equation}\label{inter4}
\begin{aligned}
\Psi^{\dagger\, s}_{\rm in}  (x+\fr{\tau}{2})
\Psi^{s}_{\rm in}(x-\fr{\tau}{2}) & \,=\,\\
 =\,\frac{e^{-i\mu\tau}}{i\tau}\,\sum_{k=0}^{\infty}(2i)^k\,&
\int\limits_{-\infty}^{\infty}
\frac{d\om_1}{\om_1}\int\limits_{\om_1}^{\infty}\frac{d\om_2}{\om_2} \cdots
\int\limits_{\om_{n-1}}^{\infty}\frac{d\om_n}{\om_n}\;
e^{i\bar{\om}x}\,\prod_{r=1}^k \sin\fr{\om_r\tau}{2}\;\\
& \hspace{4.5cm}\times \,
{\rm a}_{\sst\rm in}^{s}(\om_1)\cdots {\rm a}_{\sst\rm in}^{s}(\om_{k}),
\end{aligned}
\end{equation}
Inserting this into \rf{inter1} and noting that the integration over
$x$ produces a delta-function completes our derivation of
formula \rf{oscexpA}.

Finally we would like to show how to calculate
the leading asymptotics for $\mu\ra\infty$, $\frac{\om_r}{\mu}\ll 1$
from the general formula \rf{oscexpA}.
First, it is not hard to see that in this limit the integral
which represents $Q_{(n)}^{ss'}$, cf. equation \rf{Rdefs}, is dominated by
the contributions from small $\tau$. Approximating $\sin\frac{\om_r\tau}{2}
\simeq\frac{1}{2}\om_r\tau$ and inserting the explicit 
expression for the matrix elements
${\rm R}^{ss'}$ we arrive at
\begin{align}
& Q_{(n)}^{ss'}(\,\om_1,\dots,\om_n\,)\,\simeq \,\\ &\simeq
\,\prod_{r=1}^n(i\om_r)\int_\BR \frac{d\tau}{i\tau} \;e^{-i\tau\mu} \,
\tau^{n}\int_\BR d\om'e^{\si\frac{\pi}{2}\om'}\; e^{i\om' \tau}
\Ga\big(\fr{1}{2}+\fr{i}{2}(\om'-\om)\big)
\Ga\big(\fr{1}{2}-\fr{i}{2}(\om'+\om)\big), \nonumber\end{align}
where $\si=-$ if $s=s'$, $\si=+$ otherwise. The integral over
$\tau$ can be represented in terms of $\de(\mu-\om')$, allowing us
to do the integral over $\om'$ as well. This yields
 \begin{equation}\label{Q:appr}\begin{aligned}
Q_{(n)}^{ss'}(\,\om_1, & \dots,\om_n\,)\,\simeq \,\\& \simeq 
\,\frac{1}{i} \prod_{r=1}^n i\om_r\,\left(i\frac{\pa}{\pa\mu}\right)^{n-1}
e^{-\si\frac{\pi}{2}\mu}\,
\Ga\big(\fr{1}{2}-\fr{i}{2}(\mu+\om)\big)
\Ga\big(\fr{1}{2}+\fr{i}{2}(\mu-\om)\big).
\end{aligned}\end{equation}
With the help of Stirling's formula it is not hard to show that
\[
e^{-\si\frac{\pi}{2}\mu}
\Ga\big(\fr{1}{2}-\fr{i}{2}(\mu+\om)\big)
\Ga\big(\fr{1}{2}+\fr{i}{2}(\mu-\om)\big)\underset{\mu\ra\infty}{\simeq}
\mu^{-i\om}\left\{\begin{aligned}
1 &~{\rm if}~s=s'\\
0 &~{\rm if}~s\neq s'.
\end{aligned}\right.
\]
By inserting this relation into \rf{Q:appr} it becomes easy to
verify \rf{tree-oscexp}.

% \subsection{Proof of equation \rf{kink}}

\section{Solitonic sectors - Non-existence of normalizable ground-states}
\label{solapp}

%\subsection{}

Our aim is to prove that the sectors with nonzero fermion number
do not have normalizable ground states. The basic idea is very simple:
We should be able to decompose any state into energy eigenstates.
A state $|\Omega\rr_{n}$ could only be a
ground state in the sector with fermion number $n\neq 0$ if it would
get contributions from states with energy $-\mu$ only.
Due to the fact that the single particle spectrum is purely continuous,
one may suspect that the problem to construct normalizable
states with energy $-\mu$
is similar to the problem to construct point-like localized states
in a theory with purely continuous spectrum. There do not exist
{\it normalizable} states of this type. However, one may
be confused by the fact that we are certainly able to construct sequences
of normalized vectors which have energy expectation values that converge
to the vacuum expecation value $-\mu$. It may therefore be
worth demonstrating in some detail that no such sequence
of vectors can be convergent.

For simplicity let us restrict attention to the case in which one has
only a single set of fermionic creation- and annihilation operators
$\rmc(\om)$, $\rmc^{\dagger}(\om)$. As a preliminary remark
let us observe that the sector $\CH_f$ with fermion number $f$
may be represented as
\begin{equation}\label{enrepr}
\CH_f\,\simeq\,\int_{\BR_+}^{\oplus}d\om\;\CH_f(\om),
\end{equation}
where $\CH_f(\om)$ is generated by expressions of the form
\begin{align}
\int\limits_{-\infty}^{-\mu} & d\om_1 \dots d\om_m
  \int\limits_{-\mu}^{\infty}d\om_1'\dots d\om_{m+f}'\;
\delta\Big(\om-{\sum_{r=1}^{m+f}\om_{r}'+\sum_{s=1}^{m}\om_{s}}\Big)\\
& \times F(\om_1,\dots,\om_m;\om_1',\dots,\om_{m+f}')
\rmc(\om_1)\cdots\rmc(\om_m)\rmc^{\dagger}(\om_1')\cdots\rmc(\om_{m+f}')|\,\mu\,\rangle\!
\rangle.\nonumber
\end{align}
In the representation \rf{enrepr} one represents vectors $|\,\Psi\,\rr_f\in
\CH_f$ by vector-valued functions $\Psi_f(\om)\in\CH_f(\om)$. Upon choosing
suitable normalizations one may assume that the scalar product takes the
form
\[
\lala \,\Psi_f\,|\,\Phi_f\,\rr\,=\,\int_{-\mu}^{\infty}d\om\;
\big(\,\Psi_f(\om)\,,\,\Phi_f(\om)\,\big)_f,
\]
where $(.,.)_f$ is the scalar product in $\CH_f(\om)$.

We will consider sequences $(\Psi_n)_{n\in\BN}$ of vectors in $\CH_f$ such that
\begin{align}
{\rm (i)} &\quad \lim_{n\ra\infty}\lala\, \Psi_n\,|\,{\SH}\,|\,\Psi_n\,\rr\,=\,-\mu.
\label{convPsi}\\
{\rm (ii)} & \quad \lala\, \Psi_n\,|\,\Psi_n\,\rr\,=\,1.
\end{align}
Our aim is to show that no such sequence converges, which means that there exists
an $\epsilon>0$ such that
for any $n\in\BN$ one can find $m>n$ for which $\lVert \Psi_n-\Psi_{m}\rVert>\epsilon$.
Keeping in mind that
$\lVert \Psi_n-\Psi_{m}\rVert=2-2{\rm Re}\,\lala\Psi_n|\Psi_{m}\rr$ it
suffices to show that for any fixed $n$, the sequence
$(|\lala\Psi_n|\Psi_{m}\rr|)_{m\in\BN}$ does not converge to 1.

So let us pick any $n\in\BN$. Define $\delta>-\mu$ by
\begin{equation}\label{deltadef}
\int_{-\mu}^{\de}d\om\;\lVert\Psi_n(\om)\rVert^2_f\,=\frac{1}{2}\,.
\end{equation}
We claim that for all $\epsilon>0$ there exists $M\in\BN$ such that
for all $m>M$ we have
\begin{equation}\label{est1}
\int_{\de}^{\infty}d\om\;\lVert\Psi_{m}(\om)\rVert^2_f\,<\epsilon\,.
\end{equation}
Indeed, if this was not the case we could find an $\epsilon>0$ such that
for all $M\in\BN$ there exists $m>M$ with
$\int_{\de}^{\infty}d\om\;\lVert\Psi_m\rVert^2\,>\epsilon$, which implies that
also
\[
\begin{aligned}
\lala\, \Psi_m\,|\,{\SH}\,|\,\Psi_m\,\rr\,\geq \,&(-\mu)
\int_{-\mu}^{\de}d\om\;\lVert\Psi_m(\om)\rVert^2_f\,+\,\delta
\int_{\de}^{\infty}d\om\;\lVert\Psi_m(\om)\rVert^2_f\\
\geq \,&(-\mu)\left(1-
\int_{\de}^{\infty}d\om\;\lVert\Psi_m(\om)\rVert^2_f\right)\,+\,\delta
\int_{\de}^{\infty}d\om\;\lVert\Psi_m(\om)\rVert^2_f\\
\geq \,& -\mu+(\de+\mu)\int_{\de}^{\infty}d\om\;\lVert\Psi_m(\om)\rVert^2_f\\
> \,&-\mu+(\de+\mu)\ep.
\end{aligned}\]
Since $\de+\mu>0$ we would have a
contradiction to the convergence of the energy expecation values,
condition \rf{convPsi}.

So let us now present an estimate
for  $|\lala\Psi_n|\Psi_{m}\rr|$ that holds
for any $m$ which satisfies \rf{est1}.
\begin{align}
|\lala\Psi_n|\Psi_{m}\rr|\,
% \,=\,&
%\Bigg|
%\int\limits_{-\mu}^{\de}d\om\;\big(\Psi_n(\om),\Psi_m(\om)\big)_{\CH_n(\om)}^{}+
%\int\limits_{\de}^{\infty}d\om\;
%\big(\Psi_n(\om),\Psi_m(\om)\big)_{\CH_n(\om)}^{}\Bigg|\nonumber \\
\leq \,& \Bigg|
\int\limits_{-\mu}^{\de}d\om\;\big(\Psi_n(\om),\Psi_m(\om)\big)_{f}^{}\Bigg|+
\Bigg|\int\limits_{\de}^{\infty}d\om\;
\big(\Psi_n(\om),\Psi_m(\om)\big)_{f}^{}\Bigg|\nonumber \\
\leq \,&
\Bigg(\int\limits_{-\mu}^{\de}d\om\;\lVert\Psi_n(\om)\rVert^2_f\Bigg)^{\frac{1}{2}}
\Bigg(\int\limits_{-\mu}^{\de}d\om\;\lVert\Psi_m(\om)\rVert^2_f\Bigg)^{\frac{1}{2}}
\nonumber \\
&\qquad\qquad+
\Bigg(\int\limits_{\de}^{\infty}d\om\;\lVert\Psi_n(\om)\rVert^2_f\Bigg)^{\frac{1}{2}}
\Bigg(\int\limits_{\de}^{\infty}d\om\;\lVert\Psi_m(\om)\rVert^2_f\Bigg)^{\frac{1}{2}}
\nonumber \\
<\,&\frac{1}{2}+\frac{1}{4}\epsilon\,.\label{est2}
\end{align}
To go from the first to the second line we have used the
Cauchy-Schwartz inequality, to arrive at the last inequality we
have used \rf{deltadef} and \rf{est1}. This estimate will hold for
any $m>M$ with $M\in\BN$ being such that the validity of \rf{est1}
is guaranteed for $m>M$. Since $\ep$ is at our disposal we are
sure that $|\lala\Psi_n|\Psi_m\rr|$ will stay below $2/3<1$, say.
This clearly shows that the sequence
$(|\lala\Psi_n|\Psi_{m}\rr|)_{m\in\BN}$ can not converge to $1$.


\newpage

\begin{thebibliography}{77}

\bibitem{McGV}  J. McGreevy, H. Verlinde,
 {\it Strings from Tachyons,}
   JHEP {\bf 0312} (2003) 054

\bibitem{KMS}  I.R. Klebanov, J. Maldacena, N. Seiberg,
   {\it D-brane Decay in Two-Dimensional String Theory},
  {JHEP} {\bf 0307} (2003) 045

\bibitem{MTV} J. McGreevy, J. Teschner, H. Verlinde,
 {\it  Classical and Quantum D-branes in 2D String Theory,}
   JHEP {\bf 0401} (2004) 039

\bibitem{TT} T. Takayanagi, S. Terashima,
    {\it $c=1$ Matrix Model from String Field Theory},
   [arXiv:hep-th/0503184]

\bibitem{GR} D. Gaiotto, L. Rastelli,
  {\it A paradigm of open/closed duality: Liouville D-branes and the
       Kontsevich model} [arXiv:hep-th/0312196]

\bibitem{TT1}  T. Takayanagi, N. Toumbas,
 {\em A Matrix Model Dual of Type 0B String Theory in Two Dimensions},
 JHEP {\bf 0307} (2003) 064
 [arXiv:hep-th/0307083]

\bibitem{DKKMMS} M.R. Douglas, I.R. Klebanov, D. Kutasov, J. Maldacena,
  E. Martinec, N. Seiberg,
 {\em A New Hat For The c=1 Matrix Model},
  [arXiv:hep-th/0307195]

\bibitem{Pol}  J. Polchinski,
 {\it On the Nonperturbative Consistency of $d=2$ String Theory},
  Phys. Rev. Lett. {\bf 74} (1995) 638-641

\bibitem{SS} N. Seiberg, S. Shenker, {\it A note on
  Background (In)dependence},
  Phys. Rev. {\bf D 45} (1992) 4581-4587

\bibitem{GK} M. Gutperle, P. Kraus,
 {\it D-brane Dynamics in the c=1 Matrix Model},
  Phys. Rev. {\bf D69} (2004) 066005

\bibitem{AJ} J. Ambjorn, R.A. Janik,
 {\it The decay of quantum D-branes,}
  Phys. Lett. {\bf B584} (2004) 155-162

\bibitem{BSVY}
  J. de Boer, A. Sinkovics, E. Verlinde, J.-T. Yee,
  {\em String Interactions in c=1 Matrix Model},
 J. High-Energy Phys. {\bf 0403} (2004) 023.

%\bibitem{DFK}  P. Di Francesco, D. Kutasov,
% {\em World Sheet and Space Time Physics in Two
%  Dimensional (Super) String Theory},
%   Nucl.Phys. {\bf B375} (1992) 119-172




\bibitem{GM}  P. Ginsparg, G. Moore,
 {\it Lectures on 2D gravity and 2D string theory (TASI 1992)},
  in {\it Recent Directions in Particle Theory}, eds. J. Harvey and
  J. Polchinski, Proceedings of the 1992 TASI, World Scientific,
  Singapore, 1993

\bibitem{M} G. Moore,
 {\em Double-scaled field theory at $c=1$},
Nucl. Phys. {\bf B 368} (1992) 557-590

\bibitem{MPR} G. Moore, M.R. Plesser, S. Ramgoolam,
 {\em Exact S-matrix for two-dimensional string theory},
Nucl. Phys. {\bf B 377} (1992) 141-190

\bibitem{AKK} S. Alexandrov, I. Kostov, V. Kazakov, {\em
Time-dependent backgrounds of 2D string theory}
Nucl. Phys. {\bf B 640} (2002) 119-144; see also:
I. Kostov, {\em Integrable flows in $c=1$ string theory},
J. Phys. {\bf A36} (2003) 3153-3172

\bibitem{DRSVW} O. DeWolfe, R. Roiban, M. Spradlin, A. Volovich, J. Walcher,
 {\em On the S-matrix of Type 0 String Theory},
JHEP {\bf 0311} (2003) 012

\bibitem{Po2} J. Polchinski, {\em Classical limit of (1+1)-dimensional
   String Theory}, Nucl. Phys. {\bf B362} (1991) 125

\bibitem{MP} G. Moore, R. Plesser, {\em Classical Scattering in
   $1+1$ Dimensional String Theory}, Phys.Rev. {\bf D46} (1992) 1730-1736

\bibitem{MO} E. Martinec, K. Okuyama, {\rm Scattered Results in 2D
String Theory}, JHEP {\bf 0410} (2004) 065

\bibitem{ZZ}
A.~B.~Zamolodchikov and A.~B.~Zamolodchikov,
 {\em Liouville field theory on a pseudosphere},
[arXiv:hep-th/0101152].
%%CITATION = HEP-TH 0101152;%%


%\bibitem{FH} T. Fukuda, K. Hosomichi,
% {\em Super Liouville Theory with Boundary},
% Nucl.Phys. {\bf B635} (2002) 215-254

%\bibitem{ARS} C. Ahn, C. Rim, M. Stanishkov,
% {\em Exact One-Point Function of N=1 super-Liouville Theory with Boundary},
%Nucl.Phys. {\bf B636} (2002) 497-513

\bibitem{sen} A. Sen,  {\em Rolling tachyon},
J. High-Energy Phys. {\bf 0204} (2002) 048 ;
%%CITATION = HEP-TH 0203211;%%
 {\em Tachyon matter},
J. High-Energy Phys. {\bf 0207} (2002) 065

\bibitem{LLM}
N.~Lambert, H.~Liu and J.~Maldacena,
 {\em Closed strings from decaying D-branes},
[arXiv:hep-th/0303139].
%%CITATION = HEP-TH 0303139;%%

\bibitem{NP}  M. Natsuume, J. Polchinski,
 {\em Gravitational Scattering in the c = 1 Matrix Model}
Nucl. Phys. {\bf B424} (1994) 137-154

\bibitem{Ko} M. Kontsevich, {\em Intersection theory on the Moduli space
   of Curves and the Matrix Airy Function},
 Comm. Math. Phys. {\bf 147} (1992) 1-23


%\bibitem{BSVY}
% J. de Boer, A. Sinkovics, E. Verlinde, J.-T. Yee, {\em
% String Interactions in c=1 Matrix Model},
% J. High-Energy Phys. {\bf 0403} (2004) 023.

\bibitem{DMP} R. Dijkgraaf, G. Moore, R. Plesser: {\em
The partition function of 2D string theory},
Nucl. Phys. {\bf B394} (1993) 356-382

\bibitem{ADKMV} M. Aganagic, R. Dijkgraaf, A. Klemm, M. Marino,
C. Vafa, {\em Topological strings and integrable hierarchies},
[arXiv:hep-th/0312085]



\bibitem{AS} M. Abramowitz, I. Stegun (Eds.): Handbook of mathematical
functions (Dover, New York, 1968)

\end{thebibliography}
\end{document}